\documentclass[prd, preprint, tightenlines, a4paper, eqsecnum, nofootinbib]{revtex4-1}
\usepackage[latin1]{inputenc}
\pdfoutput=1

\usepackage{amsmath}    
\usepackage{amsfonts}
\usepackage{amssymb}
\usepackage{graphicx}   
\usepackage{color}      
\usepackage{hyperref}   
\usepackage{enumitem}   

\usepackage[textsize=tiny]{todonotes}

\allowdisplaybreaks[3]

\let\origfootnote\footnote
\renewcommand{\footnote}[1]{%
	\unskip%
	\linespread{1}%
	\origfootnote{#1}%
	\linespread{1.2}%
}

\def\bC{{\mathbb C}}       
\def\bM{{\mathbb M}}  
\def\bR{{\mathbb R}} 
\def\bS{{\mathbb S}}
\def\bH{{\mathbb H}}

\def\dd{{\mathrm{d}}}
\def\AdS{\text{AdS}}

\def\PAdS{\text{PAdS}}

\begin{document}


\author{Claudio Dappiaggi}
\email{claudio.dappiaggi@unipv.it}
\affiliation{Dipartimento di Fisica, Universit\`a degli Studi di Pavia, Via Bassi, 6, I-27100 Pavia, Italy \\
	Istituto Nazionale di Fisica Nucleare -- Sezione di Pavia, Via Bassi, 6, I-27100 Pavia, Italy}

\author{Hugo R. C. Ferreira}
\email{hugo.ferreira@pv.infn.it}
\affiliation{Istituto Nazionale di Fisica Nucleare -- Sezione di Pavia, Via Bassi, 6, I-27100 Pavia, Italy}


\title{Hadamard states for a scalar field in anti-de Sitter spacetime \\ with arbitrary boundary conditions}


\date{Revised October 2016}


\begin{abstract}
	We consider a real, massive scalar field on $\PAdS_{d+1}$, the Poincar\'e domain of the $(d+1)$-dimensional anti-de Sitter (AdS) spacetime. We first determine all admissible boundary conditions that can be applied on the conformal boundary, noting that there exist instances where ``bound states'' solutions are present. Then, we address the problem of constructing the two-point function for the ground state satisfying those boundary conditions, finding ultimately an explicit closed form. In addition, we investigate the singularities of the resulting two-point functions, showing that they are consistent with the requirement of being of Hadamard form in every globally hyperbolic subregion of $\PAdS_{d+1}$ and proposing a new definition of Hadamard states which applies to $\PAdS_{d+1}$.
\end{abstract}



\maketitle


\section{Introduction}

Quantum field theory on curved backgrounds is nowadays a well-established, thriving branch of mathematical and theoretical physics. In the past decade, especially thanks to the algebraic approach \cite{Benini:2013fia, Brunetti:2015vmh}, not only several specific models, including those with perturbative interactions, were thoroughly studied, but also foundational and structural aspects, such as renormalization or local gauge invariance, were analyzed. 

A cornerstone of most of the recent papers is the assumption that the underlying background is globally hyperbolic. This condition on the geometry of the spacetime guarantees that solutions to wave like operators, such as the Klein-Gordon, the Dirac or the Proca equation, can be found in terms of an initial value problem. As a consequence, whenever one considers a free field theory, one can follow a well-established quantization scheme, to associate with any such systems an algebra of observables, encompassing the information on structural properties such as dynamics, locality and causality. The only choice consists in the selection of a quantum state, but, also in this respect, it is nowadays universally accepted that a physically acceptable criterion lies in the so-called Hadamard condition. This is a technical requirement which guarantees on the one hand that the singular behavior of the two-point function $G^+(x,x^\prime)$ of the underlying free field theory mimics in the ultraviolet regime that of the Poincar\'e vacuum, while, on the other hand, the quantum fluctuations of all observables are finite, \cite{Kay:1988mu,Khavkine:2014mta}. As a consequence one can give a covariant definition of Wick polynomials, extending the standard one on Minkowski spacetime, and, as a by-product, interactions can be introduced at a perturbative level. In other words, if one focuses the attention on quasi-free/Gaussian states, selecting a physically acceptable state boils down to the construction of a positive, two-point function $G^+(x,x^\prime)$. This is a solution of the equation of motion in both entries, with a prescribed singular behavior. The most famous examples of Hadamard states are the Poincar\'e vacuum and the Bunch-Davies state on de Sitter spacetime \cite{Allen,BD}, but several construction schemes are nowadays known, especially on black hole \cite{Kay:1988mu,Dappiaggi:2009fx,Sanders:2013vza,Gerard} and cosmological spacetimes \cite{Olbermann:2007gn,Them:2013uka,Dappiaggi:2007mx,Dappiaggi:2008dk}.

The situation changes drastically the moment we drop the assumption of the spacetime being globally hyperbolic. Already at a classical level we face additional difficulties since we cannot construct and characterize the solutions of the equation of motion just in terms of an initial value problem. It is therefore tempting to take the easy path of concluding that these scenarios are not of interest since they bear no physical information. This attitude is not justified as there are renowned, experimentally verified effects, e.g.~the Casimir force, which are described by field theoretical models whose underlying geometry is not that of a globally hyperbolic spacetime, since the manifold possesses boundaries \cite{Dappiaggi:2014gea}.

Another relevant instance of a manifold which is not globally hyperbolic, while being central in several, important physical models is the $(d+1)$-dimensional anti de Sitter $\AdS_{d+1}$ spacetime, $d \geq 2$. This is a maximally symmetric solution of the vacuum Einstein equations with a negative cosmological constant $\Lambda$ whose underlying manifold $M\simeq\bS^1\times\bR^d$ is such that the time coordinates runs along $\bS^1$, hence yielding closed timelike curves \cite[\S 5.2]{HawkingEllis}. 

In this paper, we will focus on this class of backgrounds, more precisely on the so-called Poincar\'e fundamental domain $\PAdS_{d+1}$ which covers only a portion of the full $\AdS_{d+1}$ spacetime and which is extensively used in the prominent AdS/CFT correspondence --- see for example the recent monograph \cite{Ammon:2015wua}. Contrary to $\AdS_{d+1}$, $\PAdS_{d+1}$ can be described as the subset $\bR^d\times(0,\infty)$ of $\bR^{d+1}$ endowed with the metric $\dd s^2=\frac{\ell^2}{z^2}(-\dd t^2 + \dd z^2 + \delta^{ij} \dd x_i \dd x_j)$, $i,j=1,\ldots,d-1$, where $\ell^2=-\frac{d(d-1)}{\Lambda}$ and where $(t,z,x_i)$ are standard Cartesian coordinates with $z$ ranging only over the half line. One can realize per direct inspection that we can attach to $\PAdS_{d+1}$ a conformal, timelike boundary at $z=0$. 

From the point of view of the matter content, we will consider a real, massive scalar field, with a possibly non minimal coupling to scalar curvature. Although the dynamics is ruled by the Klein-Gordon operator, its smooth solutions cannot be constructed only starting from suitable initial data, but one needs also to prescribe boundary conditions at $z=0$. This additional input has dramatic effects at the level of quantum theory, both in the construction of the collection of all possible observables and in the identification of a physically acceptable quantum state. In this paper, we will be focusing on the second problem. As a matter of fact we will be asking ourselves two questions. The first is if one can construct the two-point function $G^+(x,x^\prime)$ for the ground state. This must be invariant under all isometries of $\PAdS_{d+1}$, a solution of the equations of motion in both entries and it must also encode the choice of boundary conditions. The second question is whether such $G^+(x,x^\prime)$ is the two-point function for a physically acceptable state. In this respect, the notion of Hadamard states cannot be invoked since it is strongly tied to spacetimes which are globally hyperbolic. When such a requirement is missing, there is no universally accepted replacement and actually this is an interesting open problem. Nonetheless, a minimal request, which one can ask, is that at least the restriction of the two-point function to any globally hyperbolic subregion of $\PAdS_{d+1}$ is of Hadamard form, a condition which can be traced back to \cite{Kay:1992es}. 

As a first step, we will be showing that for a certain range of the mass and curvature coupling no boundary condition at $z=0$ is required, whereas for its complement a whole one-parameter family of boundary conditions can be considered. As a second step, we will show that a two-point function $G^+(x,x^\prime)$ with the desired characteristics exists and it encodes in particular the choice of boundary conditions. It is important to stress that, while Dirichlet and Neumann are always unproblematic choices (whenever admissible), the Robin boundary conditions are rather tricky. As a matter of fact, we will prove that there exist instances where, upon choosing Robin boundary conditions, ``bound states'' solutions, which are exponentially suppressed for large $z$, appear. This is a very troublesome feature, first of all since it destroys the invariance under the $\PAdS_{d+1}$ isometry group. This peculiar scenario is drastically different from the usual free field theories and for this reason we will highlight its existence, leaving a more detailed analysis to future works.

In terms of the singular behavior of $G^+(x,x^\prime)$, we will show that singularities occur whenever $x$ and $x^\prime$ are connected by a null geodesic, possibly reflected at the boundary. This behavior is consistent not only with the requirement that $G^+(x,x^\prime)$ be of Hadamard form in every globally hyperbolic subregion, but also with the construction via the method of images of the two-point function, associated with the Casimir effect \cite{Dappiaggi:2014gea}, which corresponds to one of the particular cases considered here: a massless, conformally coupled scalar field.

It is important to stress that we are not the first ones to study the quantization of a real, massive scalar field in anti-de Sitter, since a first analysis appeared already in the late 1970s in \cite{Avis:1977yn}. Also the construction of a maximally symmetric two-point function was tackled before, see \cite{Allen:1985wd,Burges:1985qq}. Other recent works for this and other matter fields on AdS include \cite{Kent:2014nya,Belokogne:2016dvd}. Yet, these works considered only the special case of the Dirichlet boundary condition, which corresponds to the Friedrichs extension of the Helmholtz operator built out of the $\PAdS_{d+1}$ metric at constant time. In \cite{Ishibashi:2004wx}, the Friedrichs extension was shown to be only one of the possible self-adjoint extensions of the Helmholtz operator, which correspond to different Robin boundary conditions. In this paper, we use an alternative method to determine all these possible boundary conditions
\footnote{In the spirit of analyzing a correspondence between dynamical theories in the bulk and in the boundary of an $\AdS$ spacetime, one might wish to adapt to this case the Wentzell boundary conditions, a generalized version of the Robin ones. A preliminary, recent investigation along these lines can be found in \cite{Zahn:2015due}.} 
and, in addition, we construct the associated two-point functions for a ground state, obtaining their singular behavior.

The paper is organized as follows: In Section \ref{sec:AdS}, we will recall the basic structural, geometric properties of $\AdS_{d+1}$ and in particular of the associated Poincar\'e fundamental domain. In Section \ref{sec:KG}, we will consider the Klein-Gordon equation on $\PAdS_{d+1}$ and, by means of a conformal rescaling, we will transform it to a wave equation with a singular potential on $\mathring{\bH}^{d+1}$, the subset of Minkowski spacetime with $z>0$. After a Fourier transform in the directions orthogonal to $z$ we will reduce the dynamics to a one-dimensional ordinary differential equation of Sturm-Liouville type. This is a well studied topic, in particular with reference to the assignment of boundary conditions at the endpoint of the domain of definition of the equation. As a matter of fact, at the point $z=0$, where we want to prescribe boundary conditions, the potential of the differential equation is singular. Hence, it fails the usual idea that Dirichlet, Neumann or Robin boundary conditions are nothing but a prescription at the boundary of the behavior of a linear combination between a solution of the differential equation and its derivatives. We will outline how this obstruction can be circumvented in the language of a Sturm-Liouville problem. In Section \ref{sec:2-pt}, we will construct via a mode expansion the two-point functions for all admissible boundary conditions. We will show that for a certain class of Robin boundary conditions ``bound states'' solutions appear, while, in all other cases, one can push the analysis to the very end obtaining a closed form expression for the two-point function. In addition, we will show invariance under all isometries of the background, hence proving that we have constructed a maximally symmetric state. Finally, we will study the singular behavior of the two-point function, unveiling its consistence with the standard Hadamard prescription in all globally hyperbolic subregions and proposing a new definition of Hadamard states which apply to $\PAdS_{d+1}$.


\section{Anti-de Sitter and the Poincar\'e domain}
\label{sec:AdS}

In this paper, our starting point is the anti-de Sitter spacetime, $\AdS_{d+1}$, the maximally symmetric solution to the $(d+1)$-dimensional Einstein's equation ($d\geq 2$) with a negative cosmological constant $\Lambda$. It can be constructed starting from the embedding space $\bM^{2,d}$, that is, $\bR^{d+2}$ endowed with metric
$$\dd s^2 = \tilde{\eta}^{AB} \dd X_A \dd X_B = -\dd X^2_0 - \dd X^2_1 + \sum\limits_{i=2}^{d+1} \dd X^2_i \, , $$ 
where $(X_0,...,X_{d+1})$ are the standard Cartesian coordinates, and considering only the region identified by the relation 
\begin{equation} \label{eq:covering_space}
-X^2_0-X^2_1+\sum_{i=2}^{d+1}X^2_i=-\ell^2 \, , \qquad \ell^2 \doteq -\frac{d(d-1)}{\Lambda} \, .
\end{equation}

\begin{center}
	\begin{figure}
		\includegraphics[scale=0.425]{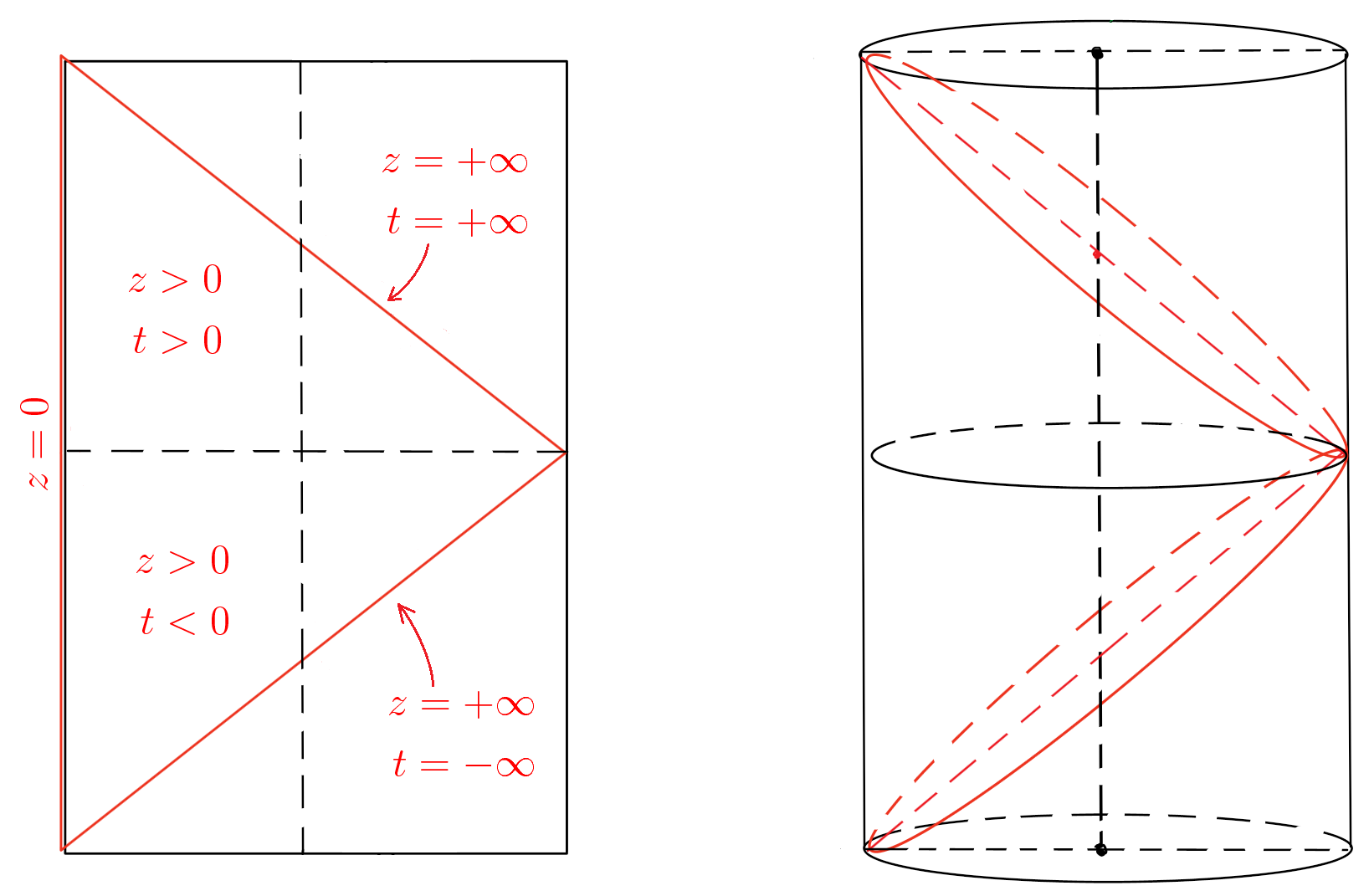}
		\caption{\label{fig:AdS}Conformal diagram of $\AdS_{d+1}$ and the Poincar\'e domain and the representation with one spatial dimension restored.}
	\end{figure}
\end{center}

For our purposes and in many physical applications, we do not work directly on $\AdS_{d+1}$, but rather on the {\em Poincar\'e fundamental domain}, $\PAdS_{d+1}$, which is identified via the coordinate transformation
\begin{equation} \label{eq:Poincare_chart}
\arraycolsep=1.4pt\def\arraystretch{2}  
\left\{\begin{array}{l}
X_0 = \dfrac{\ell}{z}t \, , \\
X_i = \dfrac{\ell}{z} x_i \, , \quad i=1,...,d-1,\\
X_d=\ell\left(\dfrac{1-z^2}{2z}+\dfrac{-t^2+\delta^{ij}x_i\,x_j}{2z}\right) \, , \\
X_{d+1}=\ell\left(\dfrac{1+z^2}{2z}-\dfrac{-t^2+\delta^{ij}x_i\,x_j}{2z}\right) \, ,
\end{array}\right.
\end{equation}
where both $t$ and all $x_i$ are ranging over the whole $\bR$, whereas $z\in (0,\infty)$. This translates the constraint which descends from the identity $X_d+X_{d+1}=\frac{\ell}{z}$, hence showing that $\PAdS_{d+1}$ covers only half of the full $\AdS_{d+1}$ (see Fig.~\ref{fig:AdS}). In addition, the metric of the Poincar\'e domain becomes
\begin{equation} \label{eq:Poincare_metric}
\dd s^2 = \frac{\ell^2}{z^2} \left(-\dd t^2+\dd z^2+\delta^{ij} \dd x_i \dd x_j\right) \, , \qquad i=1,...,d-1 \, ,
\end{equation}
where $\delta^{ij}$ stands for the Kronecker delta. Hence, $\PAdS_{d+1}$ is conformal to a portion of Minkowski spacetime, the ``upper-half plane''
$$\mathring{\bH}^{d+1}\doteq\{(t,x_1, \ldots, x_{d-1},z)\in\bR^{d+1}\;|\;z>0\} \, , $$ 
where we adopted the same Cartesian coordinates as in \eqref{eq:Poincare_metric}. If we endow $\mathring{\bH}^{d+1}$ with the standard Minkowskian metric $\eta$, then $\eta=\Omega^2 g=\frac{z^2}{\ell^2}g$ where $g$ is the metric \eqref{eq:Poincare_metric} of $\PAdS_{d+1}$ and $\Omega=\frac{z}{\ell}$ is the conformal factor.

%
%

To finish this short introduction on the geometric aspects of the background, let us briefly describe the notion of invariant distance in AdS. One can proceed in two distinct, albeit equivalent ways. Intrinsically, one can define the geodesic distance $s$ on $\PAdS_{d+1}$ between two arbitrary points $x$ and $x'$ and Synge's world function $\sigma$ given by $\sigma(x,x') \doteq \frac{1}{2}s(x,x')^2$. In view of \eqref{eq:covering_space}, one can start instead from the chordal distance $s_{\rm e}$ between $x$ and $x'$ through the embedding space $\bM^{2,d}$ and from Synge's world function defined on $\bM^{2,d}$ as
\begin{equation} \label{eq:sigmae}
\sigma_{\rm e}(x,x') \doteq \frac{1}{2}s_{\rm e}(x,x')^2 = \frac{1}{2} \tilde{\eta}^{AB} (X_A - X'_A) (X_B - X'_B) \, ,
\end{equation}
with the constraint that $x$ and $x^\prime$ are two points constrained by \eqref{eq:covering_space}, hence lying in $\AdS_{d+1}$. These two notions are related by 
\begin{equation} \label{eq:relationsigmas}
\cosh \left(\frac{s}{\ell}\right) = 1 + \frac{s_{\rm e}^2}{2 \ell^2} \, , \qquad
\cosh \left(\frac{\sqrt{2\sigma}}{\ell}\right) = 1 + \frac{\sigma_{\rm e}}{\ell^2}
\end{equation}
(see e.g.~Section 2.4 of \cite{Kent:2013}). In the rest of the paper, we set $\ell \equiv 1$.


\section{Massive scalar field on AdS}
\label{sec:KG}


\subsection{Klein-Gordon equation}

We consider a real, massive scalar field $\phi: \PAdS_{d+1}\to\bR$ such that 
\begin{equation}\label{eq:dynamicsP}
P\phi=\left(\Box_g - m_0^2 - \xi R \right)\phi=0 \, ,
\end{equation}
where $\Box_g$ is the D'Alembert wave operator built out of \eqref{eq:Poincare_metric}, $m_0$ is the mass of the scalar field, $\xi$ is the scalar-curvature coupling constant and $R = - d(d+1)$ is the Ricci scalar. 

In order to study the solutions of this equation, we follow a slightly unconventional strategy which relies on the observation made previously that $\PAdS_{d+1}$ is conformal to $\mathring{\bH}^{d+1}$ and it consists in translating \eqref{eq:dynamicsP} into a partial differential equation intrinsically defined on $\mathring{\bH}^{d+1}$. This is a standard procedure, see e.g.~Appendix D of \cite{Wald}. Let $\phi:\PAdS_{d+1}\to\bR$ be any solution of \eqref{eq:dynamicsP} and let $\Phi\doteq\Omega^{\frac{1-d}{2}}\phi$. The latter can be read as a scalar field $\Phi:\mathring{\bH}^{d+1}\to\bR$, solution of the equation
\begin{equation} \label{eq:conformally_rescaled_dynamics}
P_\eta \Phi = \left(\Box_\eta-\frac{m^2}{z^2}\right) \Phi(z)=0 \, ,
\end{equation}
in which $\Box_\eta$ is the standard wave operator built out of the Minkowski metric $\eta$ and we define
\footnote{Note that $m^2$ differs from the ``effective mass'' $m_0^2+\xi R$ used in other references.}
$m^2 \doteq m_0^2+(\xi-\frac{d-1}{4d})R$. In other words, the Klein-Gordon equation in $\PAdS_{d+1}$ is transformed to a wave equation on $\mathring{\bH}^{d+1}$ with a potential, singular at $z=0$. 

In order to construct solutions of \eqref{eq:conformally_rescaled_dynamics}, in view of the invariance of the metric under translations along the directions orthogonal to $z$, we take the Fourier transform,
\begin{equation} \label{eq:Fouriertransf}
\Phi(\underline{x},z) = \int_{\bR^d} \frac{\dd^d\underline{k}}{(2\pi)^{\frac{d}{2}}} \, e^{i\underline{k}\cdot \underline{x}} \, \widehat{\Phi}_{\underline{k}}(z) \, ,
\end{equation}
where $\underline{x} \doteq (t, x_1, \ldots, x_{d-1})$, $\underline{k} \doteq (\omega, k_1, \ldots, k_{d-1})$ and $\widehat{\Phi}_{\underline{k}}$ are solutions of
\begin{equation} \label{eq:STeq}
L \, \widehat{\Phi}_{\underline{k}} \doteq \left(- \frac{\dd^2}{\dd z^2} +\frac{m^2}{z^2} \right) \widehat{\Phi}_{\underline{k}}(z) = \lambda \, \widehat{\Phi}_{\underline{k}}(z) \, , \qquad
\lambda \equiv q^2 \doteq \omega^2 - \displaystyle\sum_{i=1}^{d-1} k_i^2 \, .
\end{equation}
This is a \emph{singular Sturm-Liouville equation}
\footnote{A good reference on singular Sturm-Liouville problems is \cite{Zettl:2005}.}
on $z \in (0,+\infty)$ with spectral parameter $\lambda$.

To fully specify a well-posed Sturm-Liouville problem, we need to add at most two boundary conditions at the endpoints 0 and $+\infty$. The required number and the form of the boundary conditions depend on the classification of the endpoints, as explained in the next section. After specifying appropriate boundary conditions, it is known that there is a continuous spectrum contained in $(0, \infty)$ and, for some boundary conditions, there is also a point spectrum with negative eigenvalues, which is indicative of the existence of ``bound states'' in the space of solutions of \eqref{eq:STeq}, that is, exponentially decaying solutions in $z$. To the best knowledge of the authors of this paper, these solutions have not been discussed so far in the literature of scalar field theory on AdS.

The next step is therefore to study the possible ($\lambda$-independent) boundary conditions that can be applied to this problem. For that purpose, as a preliminary step, we note that two linearly independent solutions of \eqref{eq:STeq} are $\sqrt{z} \, J_{\nu} \big(\sqrt{\lambda} z\big)$ and $\sqrt{z} \, Y_{\nu} \big(\sqrt{\lambda} z\big)$, 
where $J_{\nu}$ and $Y_{\nu}$ are the Bessel functions of the first and second kinds, respectively, and
\begin{equation} \label{eq:defnu}
\nu \doteq \frac{1}{2} \sqrt{1+4m^2} \, .
\end{equation}
We assume that $\nu \in [0,\infty)$ or, equivalently, $m^2 \in [-\frac{1}{4},\infty)$ (the lower bound is the Breitenlohner-Freedman bound \cite{Breitenlohner:1982jf}).


\subsection{Endpoint classification}

The types of boundary conditions that are allowed at a given endpoint depend on the integrability of the solutions near the endpoint. This classification of endpoints for a Sturm-Liouville problem has its origins with Weyl's classical limit-point and limit-circle theory \cite{Weyl}. A modern overview may be found e.g.~in \cite{Zettl:2005}. Here, we summarize the main results.

For the Sturm-Liouville problem \eqref{eq:STeq} the two endpoints are 0 and $+\infty$.

\subsubsection{Endpoint 0}

Concerning the endpoint 0, it is classified as
\begin{enumerate}[label=(\roman*)]
	\item \emph{regular} if the ``potential term'' $z \mapsto \frac{m^2}{z^2} \in L^1(0,z_0)$ for some $z_0 \in (0,+\infty)$, i.e. if $m^2=0$ ($\nu = \frac{1}{2}$); otherwise, it is \emph{singular}, i.e. if $m^2 \neq 0$ ($\nu \neq \frac{1}{2}$);
	\item \emph{limit-circle} (notation LC) if, for some $\lambda \in \bC$, all solutions of the equation are in $L^2(0,z_0)$ for some $z_0 \in (0,+\infty)$; otherwise, it is \emph{limit-point} (notation LP).
\end{enumerate}
Given that, for any $\lambda > 0$, $z \mapsto \sqrt{z} \, J_{\nu} \big(\sqrt{\lambda} z\big) \sim_{z \to 0} z^{\nu+\frac{1}{2}}$ is in $L^2(0,z_0)$ for all $\nu \in [0,\infty)$ and that $z \mapsto \sqrt{z} \, Y_{\nu} \big(\sqrt{\lambda} z\big) \sim_{z \to 0} z^{-\nu+\frac{1}{2}}$ is in $L^2(0,z_0)$ only if $\nu \in [0,1)$, for any $z_0 \in (0,+\infty)$, we may conclude that the endpoint 0 is LC for $\nu \in [0,1)$ (in particular, it is regular for $\nu = \frac{1}{2}$) and LP for $\nu \in [1,\infty)$.

\subsubsection{Endpoint $+\infty$}

As for the endpoint $+\infty$, it is classified as
\begin{enumerate}[label=(\roman*)]
	\item \emph{singular}, as it is not finite;
	\item \emph{limit-circle} (notation LC) if, for some $\lambda \in \bC$,  all solutions of the equation are in $L^2(z_0, +\infty)$ for some $z_0 \in (0,+\infty)$; otherwise, it is \emph{limit-point} (notation LP).
\end{enumerate}
For any $\lambda \in \bC$, we see that the solutions $z \mapsto \sqrt{z} \, J_{\nu} \big(\sqrt{\lambda} z\big) \sim_{z \to +\infty} \cos \big(\sqrt{\lambda} z- \frac{\nu\pi}{2} - \frac{\pi}{4}\big)$ and $z \mapsto \sqrt{z} \, Y_{\nu} \big(\sqrt{\lambda} z\big) \sim_{z \to +\infty} \sin \big(\sqrt{\lambda} z- \frac{\nu\pi}{2} - \frac{\pi}{4}\big)$ are not in $L^2(z_0, +\infty)$ for all $\nu \in [0,\infty)$. There is a solution, given by $z \mapsto \sqrt{z} \, H^{(1)}_{\nu} \big(\sqrt{\lambda} z\big) \sim_{z \to +\infty} \exp \big[i \big(\sqrt{\lambda}z- \frac{\nu\pi}{2} - \frac{\pi}{4}\big)\big]$, called the first Hankel function, which is in $L^2(z_0, +\infty)$ when ${\rm Im}(\lambda) \neq 0$, but any other linearly independent solution is not. Hence, the endpoint $+\infty$ is always LP.


\subsection{Boundary conditions}

In this section, we identify the $\lambda$-independent boundary conditions that may be assigned to the endpoints of the Sturm-Liouville problem \eqref{eq:STeq}. Their necessity and type essentially depend on the classification of the endpoints given in the previous section.

We note that in \cite{Ishibashi:2004wx} the boundary conditions that can be applied to the conformal boundary of AdS were determined by finding all self-adjoint extensions of the Helmholtz operator built out of the $\PAdS_{d+1}$ metric. Here, we give an alternative method that is consistent and complements that of \cite{Ishibashi:2004wx} and, in addition, it gives an account of the ``bound states'' solutions that occur for a class of boundary conditions.

We pick as a fundamental pair of solutions $\big\{ \widehat{\Phi}_{\underline{k}}^1, \, \widehat{\Phi}_{\underline{k}}^2 \big\}$, with
\begin{subequations}
\begin{align} 
\widehat{\Phi}_{\underline{k}}^1(z) &= \sqrt{\frac{\pi}{2}} \, q^{-\nu} \sqrt{z} \, J_{\nu}(qz) \, , \label{eq:fundamentalsolutions1} \\
\widehat{\Phi}_{\underline{k}}^2(z) &=
\begin{cases}
    - \sqrt{\dfrac{\pi}{2}} \, q^{\nu} \sqrt{z}  \, J_{-\nu}(qz) \, , & \nu \in (0,1) \, , \\
    - \sqrt{\dfrac{\pi}{2}} \sqrt{z} \left[ Y_{0}(qz) - \dfrac{2}{\pi} \log(q) \right] \, , & \nu = 0 \, .
\end{cases} \label{eq:fundamentalsolutions2}
\end{align}
\label{eq:fundamentalsolutions}
\end{subequations}

For future reference, we note that $\widehat{\Phi}_{\underline{k}}^1$ is the \emph{principal solution} at the endpoint 0, as it is the unique solution (up to scalar multiples) such that $\lim_{z \to 0^+} \widehat{\Phi}_{\underline{k}}^1(z)/\widehat{\Psi}_{\underline{k}}(z) = 0$ for every solution $\widehat{\Psi}_{\underline{k}}$ which is not a scalar multiple of $\widehat{\Phi}_{\underline{k}}^1$. The other solution $\widehat{\Phi}_{\underline{k}}^2$ is called a non-principal solution and is not unique, as it may be given by a linear combination of the principal solution with any linearly independent solution.

A general solution may then be written as 
\begin{equation}
\widehat{\Phi}_{\underline{k}}(z) = \mathcal{N}_{\underline{k}} \left[ \cos(\alpha) \, \widehat{\Phi}_{\underline{k}}^1(z) + \sin(\alpha) \, \widehat{\Phi}_{\underline{k}}^2(z) \right] \, ,
\end{equation}
where $\mathcal{N}_{\underline{k}}$ and $\alpha \in [0,\pi)$ are independent of $z$. The fundamental solutions \eqref{eq:fundamentalsolutions} were chosen such that $\alpha$ is in addition independent of $\underline{k}$.

We note that this Sturm-Liouville problem is discussed in Section~4.11 of the classical work of Titchmarsh \cite{Titchmarsh}, in the context of Fourier-Bessel expansions, and instead of $\alpha$, it is used a constant $c \in \bR$, also independent of $\underline{k}$, which is related to $\alpha$ by $c = \cot(\alpha)$.

We consider separately the following cases for different values of $\nu$.


\subsubsection{Case $\nu = \frac{1}{2}$}

This corresponds to the massless, conformally coupled scalar field. The endpoint 0 is regular in this case, whereas the endpoint $+\infty$ is singular.

The fundamental pair of solutions $\big\{ \widehat{\Phi}_{\underline{k}}^1, \, \widehat{\Phi}_{\underline{k}}^2 \big\}$ reduces to
\begin{equation}
\widehat{\Phi}_{\underline{k}}^1(z) = \sqrt{\frac{\pi z}{2q}} \, J_{\frac{1}{2}}(qz)
= \frac{\sin(qz)}{q} \, , \qquad
\widehat{\Phi}_{\underline{k}}^2(z) = - \sqrt{\frac{\pi qz}{2}} \, J_{-\frac{1}{2}}(qz)
= - \cos(qz) \, .
\end{equation}

Since the endpoint 0 is regular, the most general homogeneous boundary condition that may be applied is a \emph{Robin boundary condition} in its regular form
\begin{equation} \label{eq:BCregular}
\cos(\alpha) \, \widehat{\Phi}_{\underline{k}}(0) + \sin(\alpha) \, \widehat{\Phi}'_{\underline{k}}(0) = 0 \, , \qquad \alpha \in [0,\pi) \, .
\end{equation}
The particular case which selects the principal solution $\widehat{\Phi}_{\underline{k}}^1$, i.e.~$\alpha = 0$, is called the \emph{Friedrichs boundary condition} and it corresponds to the standard homogeneous \emph{Dirichlet boundary condition} $\widehat{\Phi}_{\underline{k}}(0) = 0$. Other common examples are the homogeneous \emph{Neumann boundary condition}, $\widehat{\Phi}'_{\underline{k}}(0) = 0$, which corresponds to $\alpha = \frac{\pi}{2}$, and the \emph{transparent boundary conditions},
\footnote{The transparent boundary conditions were used in \cite{Avis:1977yn} for the quantization of the massless, conformally coupled scalar field.}
which corresponds to $\alpha = \frac{\pi}{4}$.

An important feature occurs when we impose a Robin boundary condition with $c > 0$ or, equivalently, $\alpha \in (0, \frac{\pi}{2})$. In this case, it can be shown (Section~4.11 of \cite{Titchmarsh}) that the spectrum of the eigenvalue problem associated with the Sturm-Liouville problem \eqref{eq:STeq} does not consist purely of the continuous spectrum but it also includes a negative eigenvalue $\lambda_{\rm bs} = - c^2 = -\cot^2(\alpha)$. This indicates the existence of a ``bound state'' solution, that is, of a mode solution which exponentially decays with $z$, given by $e^{-c z} = e^{-\cot(\alpha) z}$, and
which satisfies trivially the boundary condition. This eigenvalue implies that the Fourier transform \eqref{eq:Fouriertransf} does not represent the full solution to the equation when $c > 0$. A general solution for a boundary condition of this type needs to include this ``bound state'', besides the usual propagating modes.


\subsubsection{Case $\nu \in [0,1) \setminus \{\frac{1}{2}\}$}
\label{sec:bcnusmall}

In this case, the endpoint 0 is singular and limit-circle. Hence, both solutions $\widehat{\Phi}_{\underline{k}}^1$ and $\widehat{\Phi}_{\underline{k}}^2$ are square integrable near the origin and may be used to construct a general solution. However, a Robin boundary condition written in the regular form \eqref{eq:BCregular} is no longer valid, as for instance $\lim_{z \to 0} \widehat{\Phi}_{\underline{k}}^2(z)$ diverges.

To motivate a natural way to implement a Robin boundary condition for a singular endpoint, note that in the regular case \eqref{eq:BCregular} may be equivalently written as
\begin{equation} \label{eq:BCsing}
\lim_{z \to 0} \left\{ \cos(\alpha) \, W_z \big[\widehat{\Phi}_{\underline{k}},\widehat{\Phi}_{\underline{k}}^1\big] + \sin(\alpha) \, W_z \big[\widehat{\Phi}_{\underline{k}},\widehat{\Phi}_{\underline{k}}^2\big] \right\} = 0 \, ,
\end{equation}
since $\widehat{\Phi}_{\underline{k}}^1(0) = 0$, $\big(\widehat{\Phi}_{\underline{k}}^1\big)'(0) = 1$, $\widehat{\Phi}_{\underline{k}}^2(0) = -1$ and $\big(\widehat{\Phi}_{\underline{k}}^1\big)'(0) = 0$ when $\nu = \frac{1}{2}$. In the expression, $W_z [u,v] \doteq u(z)v'(z) - v(z)u'(z)$ is the Wronskian of two differentiable functions $u$ and $v$. However, \eqref{eq:BCsing} is also valid in the singular case as the limit exists.
\footnote{In fact, the Wronskians in \eqref{eq:BCsing} are independent of $z$, but this formula remains valid if instead of the solutions $\widehat{\Phi}_{\underline{k}}^1$, $\widehat{\Phi}_{\underline{k}}^2$ we pick two functions $u$ and $v$ whose Wronskian limit is non-zero and $Lu$ and $Lv$ are square integrable near the origin (see more details in \cite{Zettl:2005}).}
Hence, one may take \eqref{eq:BCsing} as the form of a Robin boundary condition when the endpoint 0 is singular and limit-circle, a natural generalization of the regular case. Therefore, \eqref{eq:BCsing} is the most general boundary condition that can be applied for all $\nu \in (0,1)$. 

The important particular example of the Friedrichs boundary condition, which selects the principal solution at 0, corresponds to $\alpha = 0$ and we use it to define the \emph{generalized Dirichlet boundary condition}. When $\nu \in (0,1)$, we also define the \emph{generalized Neumann boundary condition} to correspond to $\alpha = \frac{\pi}{2}$, which selects the non-principal solution $\widehat{\Psi}_{\underline{k}}^2$. However, note that, given the non-uniqueness of non-principal solutions, there is no unique way to define a generalized Neumann boundary condition in the singular case which reduces to the standard definition in the regular one, $\widehat{\Phi}'_{\underline{k}}(0) = 0$. For instance, since $J_{-\frac{1}{2}}(z) = - Y_{\frac{1}{2}}(z)$, the boundary condition obtained by replacing $\widehat{\Phi}_{\underline{k}}^2$ in \eqref{eq:BCsing} by a solution proportional to $ q^{\nu} \sqrt{z} \, Y_{\nu}(qz)$ and setting $\alpha = \frac{\pi}{2}$ is not equivalent to the generalized Neumann boundary defined above.
Finally, when $\nu=0$, we define these examples of generalized boundary conditions similarly.
\footnote{The definition of Neumann boundary conditions for the non-regular cases varies from author to author, given the non-uniqueness of non-principal solutions. For instance, in \cite{Ishibashi:2004wx}, when $\nu=0$ it coincides with the Dirichlet boundary condition.}

As in the regular case, it can be shown (Section~4.11 of \cite{Titchmarsh}) that, if $c>0$, or, equivalently, $\alpha \in (0, \frac{\pi}{2})$, there is a negative eigenvalue in the spectrum of the eigenvalue problem associated with the Sturm-Liouville problem \eqref{eq:STeq}, $\lambda_{\rm bs} = - c^{1/\nu} = -\cot^{1/\nu}(\alpha)$ if $\nu \in (0,1)$ and $\lambda_{\rm bs} = - e^{-\pi c} = - e^{\pi \cot(\alpha)}$ if $\nu=0$. Hence, there is a ``bound state'' solution of the form $\sqrt{z} \, K_{\nu}(\sqrt{|\lambda_{\rm bs}|} \, z) \sim_{z \to \infty} e^{-\sqrt{|\lambda_{\rm bs}|} \, z}$, where $K_{\nu}$ is the modified Bessel solution of the second kind. This negative eigenvalue implies once more that the Fourier transform \eqref{eq:Fouriertransf} does not represent the full solution to the equation when $c > 0$. A general solution for a boundary condition of this type needs to include this ``bound state'' solution, besides the usual propagating modes.


\subsubsection{Case $\nu \in [1,\infty)$}

In this case, the endpoint 0 is singular and limit-point. Among the two fundamental solutions \eqref{eq:fundamentalsolutions}, only the principal solution $\widehat{\Phi}_{\underline{k}}^1$ is square integrable near the origin and, hence, no boundary condition is required. In practice, this is as if one had chosen the generalized Dirichlet boundary condition. Furthermore, there are no eigenvalues in the spectrum of the eigenvalue problem associated with the Sturm-Liouville problem \eqref{eq:STeq} and thus there is no ``bound state'' solution.

\vspace*{2ex}

All the cases analyzed above and the allowed boundary conditions, when necessary, are summarized in Table~\ref{tab:BCsummary}.


\setlength{\tabcolsep}{2ex}           
\renewcommand{\arraystretch}{1.25}    
\begin{center}
	\begin{table}
		\begin{tabular}{c c c}
			\hline
			$\nu = \frac{1}{2}\sqrt{1+4m^2}$ & Classification of $z=0$ & Boundary condition at $z=0$ \\
			\hline
			$\nu = \frac{1}{2}$ & Regular (R) & $\cot(\alpha) \, \widehat{\Phi}_{\underline{k}}(0) + \widehat{\Phi}_{\underline{k}}'(0) = 0$ \\
			$\nu \in [0,1), \, \nu \neq \frac{1}{2}$ & Limit-circle (LC) & $\cot(\alpha) \,  W_z\big[\widehat{\Phi}_{\underline{k}}, \widehat{\Phi}_{\underline{k}}^1 \big] + W_z\big[\widehat{\Phi}_{\underline{k}}, \widehat{\Phi}_{\underline{k}}^2 \big] = 0$ \\
			$\nu \in [1,\infty)$ & Limit-point (LP) & Not required \\
			\hline
		\end{tabular}
		\caption{Allowed boundary conditions at $z=0$, with $\alpha \in [0,\pi)$ and $\widehat{\Phi}_{\underline{k}}^1$ and $\widehat{\Phi}_{\underline{k}}^2$ defined in \eqref{eq:fundamentalsolutions}.
			\label{tab:BCsummary}}
	\end{table}
\end{center}


\section{Two-point function}
\label{sec:2-pt}

In this section we calculate the two-point or Wightman function
\footnote{In the literature of algebraic quantum field theory, the two-point function associated with a given algebraic state $\omega$ is denoted by $\omega_2$. Also, note that $G^+$ is sometimes reserved for the advanced propagator.} 
$G^+$ for a massive scalar field on $\PAdS_{d+1}$
\begin{equation}
G^+(x,x') \doteq \langle \psi | \Phi(x) \Phi(x') | \psi \rangle \, ,
\end{equation}
for the ground state $|\psi\rangle$. We perform the calculation in two ways: by a mode expansion and by closed form solutions of the differential equation satisfied by $G^+$. We show that both approaches coincide in Appendix~\ref{app:2pfcomputation}.


\subsection{Mode expansion}
\label{sec:modeexpansions}

In order to construct the two-point function for the ground state, we first use a mode expansion, a procedure already advocated in previous works, e.g.~\cite{Danielsson:1998wt}, but always in the special case of Dirichlet boundary conditions. Here, we want to consider all admissible boundary conditions discussed in the previous section. We perform the calculation for the two-point function $G^+_{\bH}$ in $\mathring{\bH}^{d+1}$, but we can immediately obtain the two-point function on $\PAdS_{d+1}$, by using the relation $G^+(x,x') = (zz')^{\frac{d-1}{2}} G^+_{\bH}(x,x')$.

Starting from \eqref{eq:conformally_rescaled_dynamics}, we look for  $G^+_{\bH}$ satisfying
\begin{equation} \label{eq:defining_Green}
\left(P_\eta\otimes\mathbb{I}\right) G^+_{\bH} = \left(\mathbb{I}\otimes P_\eta\right) G^+_{\bH} = 0 \, .  \\
\end{equation}  
where $P_\eta$ is the operator defined in \eqref{eq:conformally_rescaled_dynamics}. We consider the Fourier transform
\footnote{The Fourier transforms exists, as we are performing the computation for the ground state, which is maximally symmetric on AdS. For the two-point function of any other quantum state, it is sufficient to add a smooth, positive and symmetric bisolution of \eqref{eq:defining_Green}.}
\begin{equation} \label{eq:FouriertransG}
G^+_{\bH}(\underline{x},z;\underline{x}^\prime,z^\prime) = \int_{\bR^d} \frac{\dd^d\underline{k}}{(2\pi)^{\frac{d}{2}}} \,  e^{i\underline{k}\cdot(\underline{x}-\underline{x}^\prime)} \,  \widehat{G}^+_{\underline{k}}(z,z^\prime) \, .
\end{equation}
The remaining unknown $\widehat{G}^+_{\underline{k}}(z,z^\prime)$ is a solution of
$$ (L\otimes\mathbb{I}) \, \widehat{G}^+_{\underline{k}} = (\mathbb{I}\otimes L) \, \widehat{G}^+_{\underline{k}} = \lambda \, \widehat{G}^+_{\underline{k}} \, ,  $$
and where appropriate boundary conditions are applied at $z=0$ and $z'=0$ when $\nu \in [0,1)$.

Given that $G^+_{\bH}$ is radially symmetric in the $(d-1)$ spatial directions excluding the $z$-direction, instead of a Fourier transform along those directions we consider instead a Hankel or Fourier-Bessel transform 
\begin{gather*}
G^+_{\bH}(x,x') = \lim_{\epsilon \to 0^+} \int_{0}^{\infty} \frac{\dd \omega}{\sqrt{2\pi}} \, e^{-i\omega (t-t'-i\epsilon)}  \int_0^{\infty} \dd k \, k \, \left(\frac{k}{r}\right)^{\frac{d-3}{2}} \! J_{\frac{d-3}{2}}(kr) \, \widehat{G}^+_{\underline{k}}(z,z') \, .
\end{gather*}
where $r \doteq \sum_{i=1}^{d-1} (x^i-{x'}^i)$, only positive frequencies are taken for the ground state and $i\epsilon$ was introduced to regularize the two-point function \cite{Fulling:1987}. Finally, a change of integration variables $q^2 \doteq \omega^2 - k^2$ leads to
\begin{gather*}
G^+_{\bH}(x,x') = \lim_{\epsilon \to 0^+} \int_0^{\infty} \dd q \, q \int_0^{\infty} \dd k \, k \,\frac{e^{-i\sqrt{k^2+q^2} (t-t'-i\epsilon)}}{\sqrt{2\pi(k^2+q^2)}}   \left(\frac{k}{r}\right)^{\frac{d-3}{2}} \! J_{\frac{d-3}{2}}(kr) \, \widehat{G}^+_{\underline{k}}(z,z') \, .
\end{gather*}

At this point, some comments are in order. The \emph{antisymmetric} part of the two-point function is given by $i G(x,x')$, where $G(x,x') = \langle \psi | \left[\Phi(x), \Phi(x')\right] | \psi \rangle$ is the \emph{commutator function}. In addition to satisfying \eqref{eq:defining_Green} as the two-point function, it also satisfies
\begin{equation} \label{eq:commutatorconditions}
	G(x,x')\big|_{t=t'} = 0 \, , \qquad
	\partial_{t} G(x,x')\big|_{t=t'} = \partial_{t'} G(x,x')\big|_{t=t'} = \prod_{i=1}^{d-1} \delta(x^i-{x'}^i) \delta(z-z') \, .
\end{equation}
We can then write it as
\begin{gather*}
G(x,x') = \lim_{\epsilon \to 0^+} \sqrt{2} \int_0^{\infty} \dd q \, q \int_0^{\infty} \dd k \, k \,\frac{\sin\left(\sqrt{k^2+q^2} (t-t'-i\epsilon)\right)}{\sqrt{\pi(k^2+q^2)}}   \left(\frac{k}{r}\right)^{\frac{d-3}{2}} \! J_{\frac{d-3}{2}}(kr) \, \widehat{G}^+_{\underline{k}}(z,z') \, .
\end{gather*}
The second condition in \eqref{eq:commutatorconditions} implies that
\begin{gather*}
\sqrt{\frac{2}{\pi}} \int_0^{\infty} \dd q \, q \int_0^{\infty} \dd k \, k \, \left(\frac{k}{r}\right)^{\frac{d-3}{2}} \! J_{\frac{d-3}{2}}(kr) \, \widehat{G}^+_{\underline{k}}(z,z') = \prod_{i=1}^{d-1} \delta(x^i-{x'}^i) \delta(z-z') \, .
\end{gather*}
If we assume that $\widehat{G}_{\underline{k}}$ does not depend on $k$ (as it is the case), then by using the identity derived in Appendix~\ref{app:deltafunction}
\begin{gather*}
\int_0^{\infty} \dd k \, k \, \left(\frac{k}{r}\right)^{\frac{d-3}{2}} \! J_{\frac{d-3}{2}}(kr) 
= 2^{\frac{d-3}{2}} \, \Gamma\left(\frac{d-1}{2}\right) \, \frac{\delta(r)}{r^{d-2}} = \frac{(2\pi)^{\frac{d}{2}}\Gamma\left(\frac{d-1}{2}\right)}{\sqrt{2} \, \Gamma\left(\frac{d}{2}\right)} \prod_{i=1}^{d-1} \delta(x^i-{x'}^i) \, ,
\end{gather*}
we obtain the one-dimensional delta distribution representation
\begin{gather*}
\frac{(2\pi)^{\frac{d}{2}}\Gamma\left(\frac{d-1}{2}\right)}{\sqrt{\pi} \, \Gamma\left(\frac{d}{2}\right)} 
\int_0^{\infty} \dd q \, q \, \widehat{G}^+_{\underline{k}}(z,z') =  \delta(z-z') \, .
\end{gather*}
In other words we are looking for a resolution of the identity in terms of eigenfunctions of $L$. This problem has its roots in the theory of eigenfunction expansions and, for the case in hand, it has been tackled in Section~4.11 of \cite{Titchmarsh}. We present the results below and leave the details of their derivation to Appendix~\ref{app:eigenfunctionexpansion}.


\subsubsection{Case $\nu \in [1,\infty)$}

When $\nu \in [1,\infty)$, as we have seen in the previous section, no boundary conditions are required at the endpoint 0. The delta distribution expanded in terms of eigenfunctions of $L$ is given by
\begin{equation*}
\delta(z-z')
= \sqrt{zz'} \int_0^\infty \dd q \, q \, J_\nu(qz) J_\nu(qz') \, .
\end{equation*}
Hence, the two-point function is
\begin{equation} \label{eq:2pfnularge}
G^+_{\bH}(x,x')
= \lim_{\epsilon \to 0^+} \mathcal{N} \sqrt{zz'}
\int_0^\infty \dd k \, k \left(\frac{k}{r}\right)^{\frac{d-3}{2}} \! J_{\frac{d-3}{2}}(kr) \int_0^\infty \dd q \, q \, \frac{e^{-i\sqrt{k^2+q^2}(t-t^\prime-i\epsilon)}}{\sqrt{2\pi(k^2+q^2)}} \, J_\nu(qz) J_\nu(qz') \, ,
\end{equation}
where $\mathcal{N}$ is a normalization constant.


\subsubsection{Case $\nu \in (0,1)$}

This case, which includes the $\nu = \frac{1}{2}$ example, requires a Robin boundary condition of the form \eqref{eq:BCsing} to be applied to the solutions of the field equation. The delta distribution, expanded in terms of eigenfunctions of $L$ which satisfy the boundary condition with $c<0$, is given by
\begin{equation} \label{eq:deltaexpansionnu01}
\delta(z-z')
= \sqrt{zz'} \int_0^\infty \dd q \, q \,
\frac{\left[cJ_\nu(qz)-q^{2\nu}J_{-\nu}(qz)\right] \left[cJ_\nu(qz^\prime)-q^{2\nu}J_{-\nu}(qz^\prime)\right]}{c^2-2cq^{2\nu}\cos(\nu\pi)+q^{4\nu}} \, .
\end{equation}
Hence, for $c<0$, the two-point function is given by
\begin{align}
G^+_{\bH}(x,x')
&= \lim_{\epsilon \to 0^+} \mathcal{N} \sqrt{zz'}
\int_0^\infty \dd k \, k \left(\frac{k}{r}\right)^{\frac{d-3}{2}} \! J_{\frac{d-3}{2}}(kr) \int_0^\infty \dd q \, q \, \frac{e^{-i\sqrt{k^2+q^2}(t-t^\prime-i\epsilon)}}{\sqrt{2\pi(k^2+q^2)}} \notag \\
&\quad \times \frac{\left[cJ_\nu(qz)-q^{2\nu}J_{-\nu}(qz)\right] \left[cJ_\nu(qz^\prime)-q^{2\nu}J_{-\nu}(qz^\prime)\right]}{c^2-2cq^{2\nu}\cos(\nu\pi)+q^{4\nu}} \, .
\end{align}
If we denote $G^{+({\rm D})}_{\bH} \doteq G^+_{\bH}|_{\alpha=0}$ and $G^{+({\rm N})}_{\bH} \doteq G^+_{\bH}|_{\alpha=\frac{\pi}{2}}$, we verify that the two-point function satisfies the following boundary conditions at $z=0$ and $z'=0$
\begin{subequations} \label{eq:BCsing2pf}
	\begin{equation}
	\lim_{z \to 0} \left\{ \cos(\alpha) \, W_z \Big[G^+_{\bH},G^{+({\rm D})}_{\bH}\Big] + \sin(\alpha) \, W_z \Big[G^+_{\bH},G^{+({\rm N})}_{\bH}\Big] \right\} = 0 \, ,
	\end{equation}
	\begin{equation}
	\lim_{z' \to 0} \left\{ \cos(\alpha) \, W_{z'} \Big[G^+_{\bH},G^{+({\rm D})}_{\bH}\Big] + \sin(\alpha) \, W_{z'} \Big[G^+_{\bH},G^{+({\rm N})}_{\bH}\Big] \right\} = 0 \, .
	\end{equation}
\end{subequations}
In the particular case $\nu=\frac{1}{2}$, these reduce to
\begin{subequations}
	\begin{equation*}
	\cos(\alpha) \, G^+_{\bH}(0,z') + \sin(\alpha) \, \left.\frac{\dd G^+_{\bH}(z,z')}{\dd z}\right|_{z=0} = 0 \, ,
	\end{equation*}
	\begin{equation*}
	\cos(\alpha) \, G^+_{\bH}(z,0) + \sin(\alpha) \, \left.\frac{\dd G^+_{\bH}(z,z')}{\dd z'}\right|_{z'=0} = 0 \, .
	\end{equation*}
\end{subequations}

For $c>0$, the existence of a ``bound state'' solution with spectral parameter $\lambda = -c^{1/\nu}$ adds a contribution to the delta distribution
\begin{align*}
\delta(z-z')
&= \sqrt{zz'} \int_0^\infty \dd q \, q \,
\frac{\left[cJ_\nu(qz)-q^{2\nu}J_{-\nu}(qz)\right] \left[cJ_\nu(qz^\prime)-q^{2\nu}J_{-\nu}(qz^\prime)\right]}{c^2-2cq^{2\nu}\cos(\nu\pi)+q^{4\nu}}  \\
&\quad + 2\sqrt{zz'} \, c^{1/\nu} \, \frac{\sin(\pi \nu)}{\pi \nu} K_{\nu}\big(c^{1/(2\nu)}z\big) K_{\nu}\big(c^{1/(2\nu)}z'\big) \, .
\end{align*}
Hence, for $c>0$, the two-point function is given by
\begin{align}
G^+_{\bH}(x,x')
&= \lim_{\epsilon \to 0^+} \mathcal{N} \sqrt{zz'}
\int_0^\infty \dd k \, k \left(\frac{k}{r}\right)^{\frac{d-3}{2}} \! J_{\frac{d-3}{2}}(kr) \left\{ \int_0^\infty \dd q \, q \left[ \frac{e^{-i\sqrt{k^2+q^2}(t-t^\prime-i\epsilon)}}{\sqrt{2\pi(k^2+q^2)}} \right. \right. \notag \\
&\quad \times \left. \frac{\left[cJ_\nu(qz)-q^{2\nu}J_{-\nu}(qz)\right] \left[cJ_\nu(qz^\prime)-q^{2\nu}J_{-\nu}(qz^\prime)\right]}{c^2-2cq^{2\nu}\cos(\nu\pi)+q^{4\nu}} \right] \notag \\
&\quad + \left. 2c^{1/\nu} \, \frac{e^{-i\sqrt{k^2-c^{1/\nu}}(t-t^\prime-i\epsilon)}}{\sqrt{2\pi(k^2-c^{1/\nu})}} \, \frac{\sin(\pi \nu)}{\pi \nu} K_{\nu}\big(c^{1/(2\nu)}z\big) K_{\nu}\big(c^{1/(2\nu)}z'\big) \right\} \, .
\end{align}
The extra term is not invariant under the isometries of AdS, as it is not a function of the geodesic distance (the first term is in fact invariant, as it is shown in the next section). Therefore, it is not the two-point function for the ground state. Note, however, that it is still invariant under translations along the directions orthogonal to $z$ and $z'$, and hence the Fourier transform \eqref{eq:FouriertransG} still makes sense.


\subsubsection{Case $\nu = 0$}

This case also requires a Robin boundary condition of the form \eqref{eq:BCsing} to be applied to the solutions of the field equation. The delta distribution, expanded in terms of eigenfunctions of $L$ which satisfy the boundary condition, is given by
\begin{align*}
\delta(z-z')
&= \sqrt{zz'} \int_0^\infty \dd q \, q  
\frac{\left[(c+\frac{2}{\pi}\log(q))J_0(qz)-Y_0(qz)\right] \left[(c+\frac{2}{\pi}\log(q))J_0(qz')-Y_0(qz')\right]}{(c+\frac{2}{\pi}\log(q))^2+1} \notag \\
&\quad + 2 \sqrt{zz'} \, e^{-\pi c} \, K_0\big(e^{-\pi c/2}z\big) K_0\big(e^{-\pi c/2}z'\big) \, .
\end{align*}
For any $c \in \bR$ there is an extra contribution from a ``bound state'' solution. The two-point function is given by
\begin{align}
G^+_{\bH}(x,x')
&= \lim_{\epsilon \to 0^+} \mathcal{N} \sqrt{zz'}
\int_0^\infty \dd k \, k \left(\frac{k}{r}\right)^{\frac{d-3}{2}} \! J_{\frac{d-3}{2}}(kr) \left\{ \int_0^\infty \dd q \, q \left[ \frac{e^{-i\sqrt{k^2+q^2}(t-t^\prime-i\epsilon)}}{\sqrt{2\pi(k^2+q^2)}} \right. \right. \notag \\
&\quad \times \left. \frac{\left[(c+\frac{2}{\pi}\log(q))J_0(qz)-Y_0(qz)\right] \left[(c+\frac{2}{\pi}\log(q))J_0(qz')-Y_0(qz')\right]}{(c+\frac{2}{\pi}\log(q))^2+1} \right] \notag \\
&\quad + \left. 2e^{-\pi c} \, \frac{e^{-i\sqrt{k^2-e^{-\pi c/2}}(t-t^\prime-i\epsilon)}}{\sqrt{2\pi(k^2-e^{-\pi c/2})}} \, K_0\big(e^{-\pi c/2}z\big) K_0\big(e^{-\pi c/2}z'\big) \right\} \, . \label{eq:2pfmenuzero}
\end{align}
Similar to the case $\nu \in (0,1)$, the extra term is not invariant under the isometries of AdS, as it is not a function of the geodesic distance. Therefore, when $\nu=0$, we conclude that we are unable to construct a ground state. This may be seen as the counterpart of a massless, minimally coupled scalar field on four-dimensional de Sitter spacetime, for which there is also no ground state \cite{Kirsten:1993ug}.


\subsection{Closed form expression}
\label{sec:2pfclosedform}

The two-point function $G^+(x,x')$ for a scalar field in AdS${}_{d+1}$ on the ground state, or more generally in any maximally symmetric state, may also be given in closed form. Because of the maximal symmetry of the spacetime and of the state, it depends only on the geodesic distance between $x$ and $x'$. In Ref.~\cite{Allen:1985wd} it was shown that $G^+$ satisfies an ordinary differential equation of hypergeometric type,
\begin{equation} \label{eq:hypeq}
\left\{u(1-u) \frac{\dd^2}{\dd u^2} + \left[c-(a+b+1)u\right] \frac{\dd}{\dd u} - ab \right\} G^+(u) = 0 \, ,
\end{equation}
where
\begin{equation*}
a = \frac{d}{2} - \nu \, , \qquad
b = \frac{d}{2} + \nu \, , \qquad
c = \frac{d+1}{2} \, ,
\end{equation*}
and
\begin{equation*}
u = u(\sigma) \doteq \cosh^2 \left(\frac{\sqrt{2\sigma}}{2}\right) 
\end{equation*}
is an invariant quantity which depends only on Synge's world function  $\sigma$ defined in section~\ref{sec:AdS}. In the Poincar\'e domain, using \eqref{eq:sigmae} and \eqref{eq:relationsigmas}, $u$ may be written as
\begin{equation} \label{eq:udef}
u = \cosh^2 \left(\frac{\sqrt{2\sigma}}{2}\right) = 1 + \frac{\sigma_{\bM}}{2zz'} = \frac{\sigma_{\bM}^{(-)}}{2zz'} \, ,
\end{equation}
where, with $i=1,\ldots,d-1$,
\begin{subequations}
\begin{align*}
\sigma_{\bM} &= \frac{1}{2} \left[ - (t-t')^2 + \delta^{ij} (x_i - x'_i)(x_j - x'_j) + (z-z')^2 \right] \, , \\
\sigma_{\bM}^{(-)} &= \frac{1}{2} \left[ - (t-t')^2 + \delta^{ij} (x_i - x'_i)(x_j - x'_j) + (z+z')^2 \right] \, .
\end{align*}
\end{subequations}
Note that $u \in [0,1)$ for timelike separation and $u \in (1,\infty)$ for spacelike separation.

Two independent solutions of \eqref{eq:hypeq} when $\nu > 0$ are
\begin{subequations} \label{eq:Ghyp}
\begin{align}
G^+_1(u) &= \lim_{\epsilon \to 0^+} u_{\epsilon}^{-\frac{d}{2}-\nu} \, \frac{F \big(\tfrac{d}{2}+\nu, \tfrac{1}{2}+\nu; 1+2\nu; u_{\epsilon}^{-1}\big)}{\Gamma(1+2\nu)} \, , \label{eq:solution1} \\
G^+_2(u) &= \lim_{\epsilon \to 0^+} u_{\epsilon}^{-\frac{d}{2}+\nu} \, \frac{F \big(\tfrac{d}{2}-\nu, \tfrac{1}{2}-\nu; 1-2\nu; u_{\epsilon}^{-1}\big)}{\Gamma(1-2\nu)} \, , \label{eq:solution2}
\end{align}
\end{subequations}
where $u_{\epsilon} \doteq u(\sigma + 2i \epsilon(t-t')+\epsilon^2)$ implements the regularization of the two-point function.

The function $F(a,b;c;z)/\Gamma(c)$ (known as the regularized hypergeometric function) is an entire function of its parameters $a$, $b$ and $c$ (see e.g.~$\S 9.4$ of \cite{Lebedev:1972}). Hence, the solutions above are defined for all $\nu \geq 0$.
However, they are identical for $\nu=0$, and thus a second linearly independent solution needs to be found. In this case, two independent solutions are 
\begin{subequations} \label{eq:Ghypnu0}
\begin{align}
G^+_1(u) &= \lim_{\epsilon \to 0^+} u_{\epsilon}^{-\frac{d}{2}} \, F \big(\tfrac{d}{2}, \tfrac{1}{2}; 1; u_{\epsilon}^{-1}\big) \, , \\
G^+_2(u) &= \lim_{\epsilon \to 0^+} F \big(\tfrac{d}{2}, \tfrac{d}{2}; \tfrac{d+1}{2}; u_{\epsilon} \big) \, .
\end{align}
\end{subequations}
The second independent solution may equivalently be written as
\begin{align*}
G^+_2(u) = \lim_{\epsilon \to 0^+} \Gamma\left(\frac{d+1}{2}\right) (-u_{\epsilon})^{-\tfrac{d}{2}} 
\sum_{j=0}^{\infty} \frac{\Gamma\left(\tfrac{d}{2}+j\right) \Gamma\left(\tfrac{1}{2}+j\right)}{(j!)^2}
\left[ \log(-u_{\epsilon}) + h(j) \right] u_{\epsilon}^{-j} \, ,
\end{align*}
where
\begin{equation*}
	h(j) \doteq 2\psi(j+1) - \psi \left(\tfrac{d}{2}+j\right) - \psi \left(\tfrac{1}{2}-j\right) \, ,
\end{equation*}
and $\psi(w) \doteq \Gamma'(w)/\Gamma(w)$ is the digamma function.

These closed form expressions for the two-point functions coincide with the mode expansions obtained in the previous section. In Appendix~\ref{app:2pfcomputation}, we show that for $\nu > 0$
\begin{align*}
G^+_1(u)
&\propto \lim_{\epsilon \to 0^+} (zz')^{\frac{d}{2}}
\int_0^\infty \dd k \, k \left(\frac{k}{r}\right)^{\frac{d-3}{2}} \! J_{\frac{d-3}{2}}(kr) \int_0^\infty \dd q \, q \, \frac{e^{-i\sqrt{k^2+q^2}(t-t^\prime-i\epsilon)}}{\sqrt{2\pi(k^2+q^2)}} \, J_\nu(qz) J_\nu(qz^\prime) \, , \\
G^+_2(u)
&\propto \lim_{\epsilon \to 0^+} (zz')^{\frac{d}{2}}
\int_0^\infty \dd k \, k \left(\frac{k}{r}\right)^{\frac{d-3}{2}} \! J_{\frac{d-3}{2}}(kr) \int_0^\infty \dd q \, q \, \frac{e^{-i\sqrt{k^2+q^2}(t-t^\prime-i\epsilon)}}{\sqrt{2\pi(k^2+q^2)}} \, J_{-\nu}(qz) J_{-\nu}(qz^\prime) \, .
\end{align*}
We see that, up to normalization, $G^+_1$ is the two-point function for $\nu \in [1,\infty)$ and for $\nu \in (0,1)$ when Dirichlet boundary conditions are applied, $G^+_1 \propto G^{+({\rm D})}$, whereas $G^+_2$ is the two-point function for $\nu \in (0,1)$ when Neumann boundary conditions are applied, $G^+_2 \propto G^{+({\rm N})}$. Since they are linearly independent, we conclude that the two-point function for $\nu \in (0,1)$ and Robin boundary conditions of the form \eqref{eq:BCsing2pf} is
\begin{equation} \label{eq:2pfBCexact}
G^+(x,x') = \mathcal{N} \left[ \cos(\alpha) \, G^+_1(u) + \sin(\alpha) \, G^+_2(u) \right] \, , \qquad
\alpha \in (\tfrac{\pi}{2},\pi) \, .
\end{equation}
where $\mathcal{N}$ is a normalization constant.
For Robin boundary conditions with $\alpha \in (0,\tfrac{\pi}{2})$, the contributions from the ``bound state'' solutions obtained in Section~\ref{sec:modeexpansions} need to be added. Observe in particular that the admissible range for $\alpha$ includes also $\frac{\pi}{4}$, which corresponds to transparent boundary conditions. In this case, thus, we expect that bound states must be taken into account, a feature which was not highlighted previously in the literature.

In the case $\nu=0$, the first result in \eqref{eq:Ghypnu0} is still valid,
\begin{align*}
G^+_1(u)
&\propto \lim_{\epsilon \to 0^+} (zz')^{\frac{d}{2}}
\int_0^\infty \dd k \, k \left(\frac{k}{r}\right)^{\frac{d-3}{2}} \! J_{\frac{d-3}{2}}(kr) \int_0^\infty \dd q \, q \, \frac{e^{-i\sqrt{k^2+q^2}(t-t^\prime-i\epsilon)}}{\sqrt{2\pi(k^2+q^2)}} \, J_0(qz) J_0(qz') \, ,
\end{align*}
and thus $G^+_1$ is still the two-point function when Dirichlet boundary conditions are applied, $G^+_1 \propto G^{+({\rm D})}$, minus the contribution coming from the ``bound states''. We were unable to show explicitly that $G^+_2 \propto G^{+({\rm N})}$, but given that it is a non-principal solution of \eqref{eq:hypeq} it must be given by a linear combination of $G^{+({\rm D})}$ and $G^{+({\rm N})}$, as given in \eqref{eq:2pfmenuzero}. However, we note once more that, when $\nu=0$, the two-point function obtained above is not that of the ground state, given the lack of maximal symmetry.


\section{Hadamard condition}

In this section, we verify that the states for which the two-point functions were obtained in the previous section satisfy a natural generalization of the Hadamard condition for $\PAdS_{d+1}$.

First, recall that, for globally hyperbolic spacetimes, a quantum state is said to satisfy the \emph{local Hadamard condition} if its two-point function is of Hadamard form. A two-point function is of the Hadamard form if it is given by
\begin{align} 
	G^+(x,x') &= H^{(d+1)}(\sigma(x,x'))+ \mathcal{O}(\sigma^0) \notag \\
	&\doteq \lim_{\epsilon \to 0^+} \frac{\Gamma(\frac{d-1}{2})}{2(2\pi)^{\frac{d+1}{2}}} 
	\left[ \frac{U(x,x')}{\left(\sigma_{\epsilon}(x,x')\right)^{\frac{d-1}{2}}} + V(x,x') \log\left(\sigma_{\epsilon}(x,x')\right) + \mathcal{O}(\sigma^0) \right] \, , \label{eq:Hadamardform}
\end{align}
where $\sigma_{\epsilon} \doteq \sigma + 2i \epsilon(t-t')+\epsilon^2$ and $U$ and $V$ are smooth biscalars which are uniquely determined and only depend on the geometric features of the spacetime and on the parameter $m^2$ \cite{Decanini:2005eg}. The biscalar $V$ is identically zero for odd $d+1$ spacetime dimensions. $H^{(d+1)}(\sigma(x,x'))$ is the so-called $(d+1)$-dimensional \emph{Hadamard parametrix}. It is important to keep in mind that, having set $\ell=1$ in \eqref{eq:Poincare_metric}, $\sigma(x,x')$ is a dimensionless quantity. Hence, although in the standard version of the local Hadamard form of the two-point function the argument of the logarithm is divided by a reference scale length, in our case this is not necessary as this length has been fixed \textit{a priori}.

In globally hyperbolic spacetimes, it follows that the local Hadamard condition is equivalent to the global Hadamard one \cite{Radzikowski:1996pa,Radzikowski:1996ei}, which entails in addition that the only singularity of the two-point function is at $\sigma=0$ and that it is of Hadamard form. A more rigorous definition requires the tools of microlocal analysis and may be found in \cite[Ch.5]{Brunetti:2015vmh}.

Even though AdS is not globally hyperbolic, we can still verify if the two-point functions obtained above are of Hadamard form for every globally hyperbolic subregion. If that is the case, we say that the maximally symmetric state in AdS satisfies the local Hadamard condition. However, it does not follow that the state satisfies a global Hadamard condition, as the standard definition, adopted in globally hyperbolic spacetimes, does not apply. A novel analysis is required and we plan to address it in future work \cite{DappiaggiFerreira}, also in view of the investigation in \cite{Vasy}. Here, we verify that the two-point functions in $\PAdS$ have a richer singularity structure
\footnote{As a side comment, this feature resembles what happens in de Sitter spacetime when one considers the so-called $\alpha$-vacua \cite{Brunetti:2005pr}, although, in this case, the additional singularities are pathological, being the underlying background globally hyperbolic.}
than those in globally hyperbolic spacetimes, while at the same time satisfying the local Hadamard condition in any globally hyperbolic subregion.

We will focus on the study of the singularities of the two-point functions obtained for the ground state in the cases of $d=2$ and $d=3$. Analogous comments can be made for larger $d$, as we discuss briefly below. 

We start with $d=3$, the physically relevant case, and assume $\nu > 0$. The two-point function is a linear combination of the solutions \eqref{eq:Ghyp} and we know that the hypergeometric functions in those solutions have only three singular points: $u=0, \, 1, \infty$. The latter, $u \to \infty$, occurs when either $z \to 0$ or $z' \to 0$, which takes one of the points $x, \, x'$ to the boundary and, therefore, does not belong to the spacetime.

The singularity $u=1$ corresponds to $\sigma = 0$, cf.~\eqref{eq:udef}. If we expand the solutions \eqref{eq:Ghyp} with $d=3$ in $\sigma$, such that $x$ and $x'$ belong to a globally hyperbolic subregion of $\AdS_4$,
\begin{subequations}
\begin{align*}
G^+_1(u_{\epsilon}) &= \frac{2^{1+2\nu} \, \Gamma \left(1+\nu\right)}{\sqrt{\pi} \, \Gamma \left(\frac{3}{2}+\nu\right) \Gamma \left(1+2\nu\right)} \left[ \frac{1}{\sigma_{\epsilon}} + \frac{1}{2} \left(\nu^2 - \frac{1}{4}\right) \log(\sigma_{\epsilon}) + \mathcal{O}(\sigma^0) \right] \, , \\
G^+_2(u_{\epsilon}) &= \frac{2^{1-2\nu} \, \Gamma \left(1-\nu\right)}{\sqrt{\pi} \, \Gamma \left(\frac{3}{2}-\nu\right) \Gamma \left(1-2\nu\right)} \left[ \frac{1}{\sigma_{\epsilon}} + \frac{1}{2} \left(\nu^2 - \frac{1}{4}\right) \log(\sigma_{\epsilon}) + \mathcal{O}(\sigma^0) \right] \, .
\end{align*}
\end{subequations}
This expansion is exactly the Hadamard expansion \eqref{eq:Hadamardform}, up to normalization constants, presented for $d=3$ in \cite{Decanini:2005eg} for a globally hyperbolic subregion of $\AdS_4$. Hence, in view of \eqref{eq:2pfBCexact}, the two-point function reads
\begin{equation*}
G^+(x,x') \propto \left[\cos(\alpha) + \sin(\alpha)\right] H^{(4)}(\sigma(x,x'))+\mathcal{O}(\sigma^0) \, .
\end{equation*}
By choosing a suitable $\alpha$-dependent normalization constant, we can make $G^+(x,x')$ satisfy the local Hadamard property \eqref{eq:Hadamardform}, except if $\alpha = \frac{3\pi}{4}$.

\begin{figure}
	\begin{center}
		\includegraphics[scale=1]{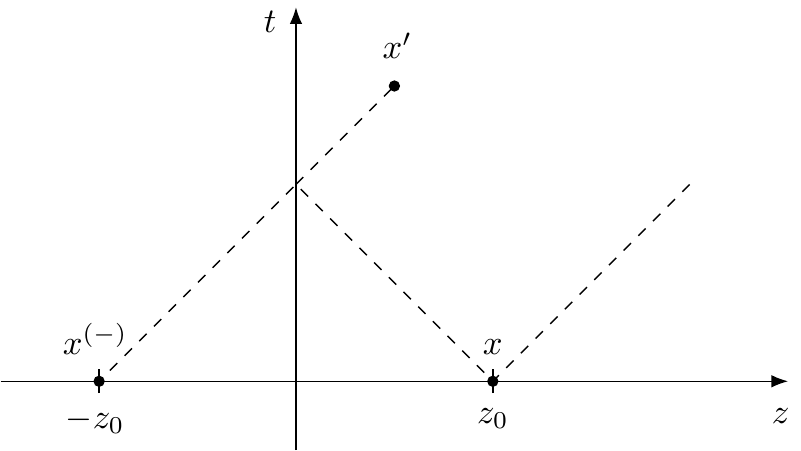}
	\end{center}
	\caption{\label{fig:wavefrontset}Singularity structure of the two-point function.}
\end{figure}

The singularity $u=0$ corresponds to $\sigma^{(-)} = 0$, where $\sigma^{(-)}$ is such that, cf.~Eq.~\eqref{eq:udef},
$$u \doteq - \sinh^2 \left(\frac{\sqrt{2\sigma^{(-)}}}{2} \right) = \frac{\sigma_{\bM}^{(-)}}{2zz'} \, . $$
Two points $x$ and $x'$ are such that $\sigma^{(-)}(x,x') = 0$ if there is a null geodesic starting at $x$ that is ``reflected'' at the boundary and ends at $x'$ (see Fig.~\ref{fig:wavefrontset}). More rigorously, if we consider the conformally related spacetime $\mathring{\bH}^4$ and allow $z$ to take all real values, $\sigma^{(-)}(x,x')$ vanishes if $\sigma_{\bM}^{(-)}(x,x') = 0$, or equivalently if $\sigma_{\bM}(x^{(-)},x') = 0$, where $x^{(-)} \doteq x|_{z \mapsto -z}$. Note that there is no globally hyperbolic subregion of $\AdS_4$ in which $\sigma^{(-)}(x,x') = 0$, hence this singularity is not present for a two-point function on a globally hyperbolic submanifold.


\begingroup\abovedisplayskip=9pt \belowdisplayskip=9pt

If we now expand the two solutions in $\sigma^{(-)}$, we obtain
\begin{subequations}
	\begin{align*}
	G^+_1(u_{\epsilon}) &= i (-1)^{\nu} \frac{2^{1+2\nu} \, \Gamma \left(1+\nu\right)}{\sqrt{\pi} \, \Gamma \left(\frac{3}{2}+\nu\right) \Gamma \left(1+2\nu\right)} \left[\frac{1}{\sigma_{\epsilon}^{(-)}} + \frac{1}{2} \left(\nu^2 - \frac{1}{4}\right) \log(\sigma_{\epsilon}^{(-)}) + \mathcal{O}\big((\sigma^{(-)})^0\big) \right] \, , \\
	G^+_2(u_{\epsilon}) &= i (-1)^{-\nu} \frac{2^{1-2\nu} \, \Gamma \left(1-\nu\right)}{\sqrt{\pi} \, \Gamma \left(\frac{3}{2}-\nu\right) \Gamma \left(1-2\nu\right)} \left[\frac{1}{\sigma_{\epsilon}^{(-)}} + \frac{1}{2} \left(\nu^2 - \frac{1}{4}\right) \log(\sigma_{\epsilon}^{(-)}) + \mathcal{O}\big((\sigma^{(-)})^0\big) \right] \, .
	\end{align*}
\end{subequations}
This has exactly the same Hadamard form, up to normalization constants, but with respect to $\sigma^{(-)}$. Hence, in view of \eqref{eq:2pfBCexact}, the two-point function reads
\begin{equation*}
G^+(x,x') \propto \left[\cos(\alpha) + (-1)^{-2\nu} \sin(\alpha)\right] H^{(4)}(\sigma^{(-)}(x,x'))+\mathcal{O}\big((\sigma^{(-)})^0\big) \, .
\end{equation*}
The singular contribution vanishes for $\nu=\frac{1}{2}$ and $\alpha = \frac{\pi}{4}$, for which there are no singularities along reflected null geodesics. This justifies why $\alpha = \frac{\pi}{4}$ is referred to as transparent boundary conditions for the massless, conformally coupled scalar field.

We analyze now the two-point function for $d=2$ and $\nu > 0$. The two solutions \eqref{eq:Ghyp} with $d=2$ have the same two singularities at $u=1$ and $u=0$. If we expand them in $\sigma$, such that $x$ and $x'$ belong to a globally hyperbolic subregion of $\AdS_3$, we obtain
\begin{equation*}
G^+_1(u_{\epsilon}) = \frac{2^{\frac{1}{2}+2\nu}}{\Gamma \left(1+2\nu\right)} \frac{1}{\sqrt{\sigma_{\epsilon}}} + \mathcal{O}(\sigma^0) \, , \qquad
G^+_2(u_{\epsilon}) = \frac{2^{\frac{1}{2}-2\nu}}{\Gamma \left(1-2\nu\right)} \frac{1}{\sqrt{\sigma_{\epsilon}}} + \mathcal{O}(\sigma^0) \, .
\end{equation*}
Again, this is of the same Hadamard form \eqref{eq:Hadamardform} as presented in \cite{Decanini:2005eg} for $d=2$ in any globally hyperbolic subregion of $\AdS_3$. Hence, in view of \eqref{eq:2pfBCexact}, the two-point function reads
\begin{equation*}
G^+(x,x') \propto \left[\cos(\alpha) + \sin(\alpha)\right] H^{(3)}(\sigma(x,x'))+\mathcal{O}(\sigma^0) \, .
\end{equation*}
If we instead expand them in $\sigma^{(-)}$, we obtain
\begin{equation*}
G^+_1(u_{\epsilon}) = i (-1)^{\nu} \frac{2^{\frac{1}{2}+2\nu}}{\Gamma \left(1+2\nu\right)} \frac{1}{\sqrt{\sigma^{(-)}_{\epsilon}}} + \mathcal{O}(\sigma^0) \, , \quad
G^+_2(u_{\epsilon}) = i (-1)^{-\nu} \frac{2^{\frac{1}{2}+2\nu}}{\Gamma \left(1+2\nu\right)} \frac{1}{\sqrt{\sigma^{(-)}_{\epsilon}}} + \mathcal{O}(\sigma^0) \, .
\end{equation*}
Hence, in view of \eqref{eq:2pfBCexact}, the two-point function reads
\begin{equation*}
G^+(x,x') \propto i (-1)^{\nu} \left[\cos(\alpha) + (-1)^{-2\nu} \sin(\alpha)\right] H^{(3)}(\sigma^{(-)}(x,x'))+\mathcal{O}\big((\sigma^{(-)})^0\big) \, .
\end{equation*}
Therefore, we shall also call a quantum state \emph{Hadamard} in $\PAdS_3$ a state whose two-point function has the singularity structure described above.

\endgroup

Similar investigations can be made for larger $d$. However, it becomes increasingly impractical to perform the expansions in $\sigma$ and $\sigma^{(-)}$ since we would have to resort to a case by case analysis. There are recursive methods to obtain the expansions of $U$ and $V$ in $\sigma$ for any fixed $d$, but they get significantly more complex for larger $d$ (detailed expressions for $d+1$ up to 6 may be found in Ref.~\cite{Decanini:2005eg}). Nevertheless, using the tools of microlocal analysis, it is possible to show that the singularity structure observed for $d=2$ and $d=3$, with which we defined the notion of a Hamadard state on $\PAdS_3$ and $\PAdS_4$, is verified for any $d$. We leave this proof to a forthcoming work \cite{DappiaggiFerreira}. 

In view of the above analysis, we define a \emph{Hadamard} quantum state in $\PAdS_{d+1}$, $d \geq 2$, to be any state whose two-point function $G^+(x,x')$ is such that 
$$G^+(x,x') - H^{(d+1)}(\sigma(x,x')) - i (-1)^{-\nu} \frac{\cos(\alpha) + (-1)^{-2\nu} \sin(\alpha)}{\cos(\alpha) + \sin(\alpha)} \, H^{(d+1)}(\sigma^{(-)}(x,x')) $$
is a smooth function on $\PAdS_{d+1}\times\PAdS_{d+1}$. In particular, if $\alpha = \frac{3\pi}{4}$, we cannot find a Hadamard state satisfying this definition.

Notice that although a ground state does not exist for $\alpha \in (0,\frac{\pi}{2})$, on account of the presence of ``bound state'' solutions, the proposed definition still applies to these cases.


\section{Conclusions}  

In this paper, we have considered a real, massive scalar field on $\PAdS_{d+1}$, the Poincar\'e domain of the $(d+1)$-dimensional AdS spacetime. In particular, we have determined all admissible boundary conditions that can be applied on the conformal boundary and we have constructed the two-point function associated with the ground state, finding ultimately an explicit closed form. In addition, we have investigated its singular structure, showing consistency with the minimal requirement of being of Hadamard form in every globally hyperbolic subregion of $\PAdS_{d+1}$. As a consequence we propose a new definition of Hadamard states which applies to $\PAdS_{d+1}$.

To conclude our work, we would like to highlight two open issues which we deem appropriate of further investigations. The first concerns the choice of boundary conditions. As we have shown, there are instances where ``bound state'' solutions appear in the construction of the two-point and of the commutator functions. The main direct consequence of this unexpected feature is the lack of a ground state for the underlying system, as invariance under the action of certain isometries is broken. On the one hand, we can observe that this poses no obstruction to the existence of Hadamard states, but, on the other hand, there is no clear physical interpretation why such ``bound state'' solutions appear and what are the concrete consequences of their existence.

The second open problem lies in the investigation of the notion of Hadamard states for a real, massive scalar field on $\PAdS_{d+1}$. In the last section we have given a local definition, which exploits ultimately the existence of a global coordinate chart on $\mathring{\bH}^{d+1}$. If one aims at generalizing these results to asymptotically AdS spacetimes or even to manifolds with timelike boundaries, we cannot expect that a local construction becomes practical. Hence, following a similar path to the one taken by those who investigated Hadamard states on globally hyperbolic spacetimes, we expect that a necessary step is to translate our analysis in the language of microlocal analysis. In this way, we hope to be able to give a global definition of Hadamard states and to formulate a version of the work of Radzikowski \cite{Radzikowski:1996ei,Radzikowski:1996pa} in the context of asymptotically AdS spacetimes.


\section*{Acknowledgments}  
We are grateful to Jorma Louko for enlightening discussions and for pointing out Ref.~\cite{Titchmarsh}. We are also grateful to Nicol\`o Drago, Gabriele Nosari, Pedro Lauridsen Ribeiro, Nicola Pinamonti and Micha{\l} Wrochna for useful comments and discussions. The work of C.D. was supported by the University of Pavia. The work of H.~F. was supported by the INFN postdoctoral fellowship ``Geometrical Methods in Quantum Field Theories and Applications''.


\appendix


\section{Two-point function computation for $\nu>0$}
\label{app:2pfcomputation}

The two-point function for a massive scalar field in $\mathring{\bH}^{d+1}$ for $\nu \in [1,\infty)$ or in the case of Dirichlet boundary conditions for $\nu \in (0,1)$ is given by \eqref{eq:2pfnularge},
\begin{equation} \label{eq:tpf1}
G^+_{\bH}(x,x')
= \mathcal{N} \sqrt{zz'} \int_0^\infty \dd k \, k \left(\frac{k}{r}\right)^{\frac{d-3}{2}} \! J_{\frac{d-3}{2}}(kr) \int_0^\infty \dd q \, q \, \frac{e^{-i\sqrt{k^2+q^2}(t-t^\prime-i\epsilon)}}{\sqrt{2\pi(k^2+q^2)}} \, J_\nu(qz) J_\nu(qz') \, ,
\end{equation}
where we omit the limit $\epsilon \to 0^+$ for presentation simplicity. The two-point function in the case of Neumann boundary conditions when $\nu \in (0,1)$ can also be obtained from \eqref{eq:tpf1} by allowing $\nu \in (-1,0)$ (see section~\ref{sec:bcnusmall}). 

In this appendix, we compute explicitly the integrals in \eqref{eq:tpf1} and obtain the two-point function in closed form, as presented in Section~\ref{sec:2pfclosedform}.

Using Eqs.~(6.737.5) and (6.737.6) of \cite{Gradshteyn}, for $d=2, \, 3$, we obtain
\begin{align}
G^+_{\bH}(x,x')
&= \mathcal{N} \sqrt{zz'} \int_0^\infty \dd q \, q^{\frac{d}{2}} \, J_{\nu}(qz) J_\nu(qz^\prime) 
\left\{ \frac{1}{\pi} \, \Theta(r-(t-t')) \frac{K_{\frac{d}{2}-1} \left(q  \sqrt{\chi_{\epsilon}^2}\right)}{\left(\sqrt{\chi_{\epsilon}^2}\right)^{\frac{d}{2}-1}} \right. \notag \\
&\quad \left. -\frac{i}{2} \Theta(t-t'-r) \frac{J_{1-\frac{d}{2}} \left(q \sqrt{-\chi_{\epsilon}^2}\right) - i \, Y_{1-\frac{d}{2}} \left(q \sqrt{-\chi_{\epsilon}^2}\right)}{\left(\sqrt{-\chi_{\epsilon}^2}\right)^{\frac{d}{2}-1}} \right\} \notag \\
&= \mathcal{N} \, \frac{\sqrt{zz'}}{\pi} \int_0^\infty \dd q \, q^{\frac{d}{2}} \, \frac{K_{\frac{d}{2}-1} \left(q  \sqrt{\chi_{\epsilon}^2}\right)}{\left(\sqrt{\chi_{\epsilon}^2}\right)^{\frac{d}{2}-1}} \, J_{\nu}(qz)J_\nu(qz^\prime)  \, , \label{eq:2pfintegralstep1}
\end{align}
where $\chi_{\epsilon}^2 \doteq r^2-(t-t'-i\epsilon)^2$, $\Theta$ is the Heaviside function, $K_{\frac{d}{2}-1}$ is the modified Bessel function of the second kind and we used (see e.g.~\cite[\S 5.6-5.7]{Lebedev:1972})
\begin{gather*}
	J_{\alpha}(w) - i Y_{\alpha}(w) = H_{\alpha}^{(2)}(w) = e^{i\pi\alpha} H_{-\alpha}^{(2)}(w) \, , 
	\qquad w \notin (-\infty,0] \, , \\
	K_{\alpha}(w) = - \frac{i\pi}{2} e^{-\frac{i\pi\alpha}{2}} H_{\alpha}^{(2)}(-iw) \, ,
	\qquad \arg(w) \in \left[-\frac{\pi}{2},\pi\right] \, ,
\end{gather*}
where $H_{\alpha}^{(2)}$ is the second Hankel function. Even though the calculation leading to \eqref{eq:2pfintegralstep1} is valid for $d=2, \, 3$, the result can be analytically continued to $d \geq 2$.


At this point, it is convenient to consider even and odd $d$ separately. Let $d = 2n+1$, $n=1, 2, \ldots$. Then,
\begin{align}
G^+_{\bH}(x,x')
&= \mathcal{N} \, \frac{\sqrt{zz'}}{\pi} \int_0^\infty \dd q \, q^{n+\frac{1}{2}} \, \frac{K_{n-\frac{1}{2}} \left(q  \chi_{\epsilon}\right)}{\chi_{\epsilon}^{n-\frac{1}{2}}} \, J_{\nu}(qz)J_\nu(qz') \notag \\
&= \mathcal{N} \, \frac{\sqrt{zz'}}{\pi} \left. \left(\frac{1}{\chi} \frac{\dd}{\dd \chi} \right)^n \int_0^\infty \dd q \, \frac{K_{-\frac{1}{2}} \left(q  \chi\right)}{\left(q\chi\right)^{-\frac{1}{2}}} \, J_{\nu}(qz)J_\nu(qz') \right|_{\chi=\chi_{\epsilon}} \notag \\
&= \mathcal{N} \, \frac{\sqrt{zz'}}{\sqrt{2\pi}} \left. \left(\frac{1}{\chi} \frac{\dd}{\dd \chi} \right)^n \int_0^\infty \dd q \, e^{-q \chi} \, J_{\nu}(qz)J_\nu(qz') \right|_{\chi=\chi_{\epsilon}} \notag \\
&= \mathcal{N} \, \frac{1}{\sqrt{2\pi^3}} \left. \left(\frac{1}{\chi} \frac{\dd}{\dd \chi} \right)^n Q_{\nu-\frac{1}{2}}\left( \frac{z^2+{z'}^2+\chi^2}{2zz'} \right) \right|_{\chi=\chi_{\epsilon}} \notag \\
&= \mathcal{N} \, \frac{1}{\sqrt{2\pi^3}(zz')^n} \, \left. \frac{\dd^{n}}{\dd \eta^{n}}  Q_{\nu-\frac{1}{2}}\left(\eta\right) \right|_{\eta=\eta_{\epsilon}} \notag \\
&= \mathcal{N} \, \frac{1}{\sqrt{2\pi^3}(2zz')^n} \, \left. \frac{\dd^{n}}{\dd u^{n}}  Q_{\nu-\frac{1}{2}}\left(2u-1\right) \right|_{u=u_{\epsilon}} \notag \\
&= \mathcal{N} \, \frac{1}{\sqrt{2\pi^3}(zz')^n} \, \left. \left(\frac{1}{\sinh(s)} \frac{\dd}{\dd s}\right)^n Q_{\nu-\frac{1}{2}}\left(\cosh(s)\right) \right|_{s=s_{\epsilon}} \, , \label{eq:2pfQ}
\end{align}
where 
\begin{equation*} 
\eta_{\epsilon} \doteq \frac{z^2+{z'}^2+r^2-(t-t'-i\epsilon)^2}{2zz'} = 2u_{\epsilon} - 1 \doteq \cosh(s_{\epsilon}) \, ,
\end{equation*}
$Q_{\nu-\frac{1}{2}}$ is the Legendre function of the second kind and where we used Eq.~(6.612.3) of \cite{Gradshteyn} and the relation
\begin{equation} \label{eq:derbesselK}
	\left(\frac{1}{\chi} \frac{\dd}{\dd \chi} \right)^n \left( \chi^{-\alpha} K_{\alpha}(\chi) \right) = \chi^{-\alpha-n} K_{\alpha+n}(\chi) \, .
\end{equation}

Note that \eqref{eq:2pfQ} is valid for $\nu > -\frac{1}{2}$ but is not defined for $\nu = -1/2$, hence it cannot be used as currently written for the case of Neumann boundary conditions. However, we can extend it analytically to $\nu > -1$ as follows. From Eq.~(14.10.4) of \cite{NIST},
\begin{equation} \label{eq:Qderrec}
(1-\eta_{\epsilon}^2) \, Q'_{\nu-\frac{1}{2}}(\eta_{\epsilon}) = \left(\nu+\frac{1}{2}\right) \left[\eta_{\epsilon} \, Q_{\nu-\frac{1}{2}}(\eta_{\epsilon}) - Q_{\nu+\frac{1}{2}}(\eta_{\epsilon}) \right] \, .
\end{equation}
$Q_{\nu-\frac{1}{2}}$ is not defined for $\nu = -1/2$, as for Eq.~(14.3.7) of \cite{NIST} one has
\begin{equation*}
Q_{\nu-\frac{1}{2}}(\eta) = \frac{\sqrt{\pi} \, \Gamma\left(\nu+\frac{1}{2}\right)}{2^{\nu-\frac{1}{2}} \, \eta^{\nu+\frac{1}{2}} \, \Gamma(\nu+1)} \, F\left(\tfrac{\nu}{2}+\tfrac{3}{4}, \tfrac{\nu}{2}+\tfrac{1}{4}; \nu+1; \tfrac{1}{\eta^2}\right) \, ,
\end{equation*}
for $\nu \notin -\frac{2\mathbb{N}+1}{2}$ and $\eta>1$. Nevertheless, one can analytically continue \eqref{eq:Qderrec} to $\nu > -1$ as
\begin{equation*}
(1-\eta_{\epsilon}^2) \, Q'_{\nu-\frac{1}{2}}(\eta_{\epsilon}) = \frac{\sqrt{\pi} \, \Gamma\left(\nu+\frac{3}{2}\right)}{2^{\nu-\frac{1}{2}} \, \eta_{\epsilon}^{\nu-\frac{1}{2}} \, \Gamma(\nu+1)} \, F\left(\tfrac{\nu}{2}+\tfrac{3}{4}, \tfrac{\nu}{2}+\tfrac{1}{4}; \nu+1; \tfrac{1}{\eta_{\epsilon}^2}\right) - \left(\nu+\frac{1}{2}\right) Q_{\nu+\frac{1}{2}}(\eta_{\epsilon}) \, .
\end{equation*}
Using the same notation for the extended function, \eqref{eq:2pfQ} may be used for the Neumann boundary conditions with $\nu \in (-1,0)$.


Let $d = 2n$, $n=1, 2, \ldots$. Then,
\begin{align*}
G^+_{\bH}(x,x')
&= \mathcal{N} \, \frac{\sqrt{zz'}}{\pi} \int_0^\infty \dd q \, q^{n} \, \frac{K_{n-1} \left(q  \chi_{\epsilon}\right)}{\chi_{\epsilon}^{n-1}} \, J_{\nu}(qz)J_\nu(qz') \\
&= \mathcal{N} \, \frac{\sqrt{zz'}}{\pi} \left. \left(\frac{1}{\chi} \frac{\dd}{\dd \chi} \right)^{n-1} \int_0^\infty \dd q \, q \, K_0 \left(q  \chi\right) J_{\nu}(qz)J_\nu(qz') \right|_{\chi=\chi_{\epsilon}} 
\\
&= \mathcal{N} \, \frac{\sqrt{zz'}}{\pi} \left. \left(\frac{1}{\chi} \frac{\dd}{\dd \chi} \right)^{n-1}  \frac{\left(\frac{\sqrt{\chi^2+(z+z')^2}+\sqrt{\chi^2+(z-z')^2}}{\sqrt{\chi^2+(z+z')^2}-\sqrt{\chi^2+(z-z')^2}}\right)^{-\nu}}{\sqrt{(\chi^2+(z+z')^2)(\chi^2+(z-z')^2)}}  \right|_{\chi=\chi_{\epsilon}} \\
&= \mathcal{N} \, \frac{2^{\nu-1}}{\pi(zz')^{n-\frac{1}{2}}} \left. \frac{\dd^{n-1}}{\dd \eta^{n-1}}  \frac{\left(\eta+\sqrt{\eta^2-1}\right)^{-\nu}}{\sqrt{\eta^2-1}}  \right|_{\eta=\eta_{\epsilon}}  \\
&= \mathcal{N} \, \frac{2^{\nu-2}}{\pi(zz')^{n+\frac{1}{2}}} \left. \frac{\dd^{n-1}}{\dd u^{n-1}}  \frac{\left(2u-1+\sqrt{u(u-1)}\right)^{-\nu}}{\sqrt{u(u-1)}}  \right|_{u=u_{\epsilon}}  \\
&= \mathcal{N} \, \frac{2^{\nu-1}}{\pi(zz')^{n-\frac{1}{2}}} \left. \left(\frac{1}{\sinh(s)} \frac{\dd}{\dd s}\right)^{n-1} \frac{e^{-\nu s}}{\sinh(s)} \right|_{s=s_{\epsilon}} \, ,
\end{align*}
where we used Eq.~(6.522.3) of \cite{Gradshteyn} and \eqref{eq:derbesselK}.


To prove that these results are equivalent to the ones written in terms of the hypergeometric functions \eqref{eq:Ghyp}, we show that they satisfy the same initial conditions, since they are all solutions of the same differential equation.

First, we verify the claim for $d=2, \, 3$. In terms of the invariant quantity $u$, for $d=2$, let
\begin{align*}
	g_1^{d=2}(u) &= u_{\epsilon}^{-1-\nu} F\left(1+\nu, \tfrac{1}{2}+\nu; 1+2\nu; u_{\epsilon}^{-1} \right) \, , \\
	g_2^{d=2}(u) &= 4^{\nu} \, \frac{\left(2u_{\epsilon}-1+2\sqrt{u_{\epsilon}(u_{\epsilon}-1)}\right)^{-\nu}}{\sqrt{u_{\epsilon}(u_{\epsilon}-1)}} \, .
\end{align*}
Then,
\begin{equation*}
	g_1^{d=2}(u) = g_2^{d=2}(u) = u_{\epsilon}^{-1-\nu} \left(1 + \frac{1+\nu}{2u} + \mathcal{O}(u^{-2}) \right) \, .
\end{equation*}
Hence, $g_1^{d=2} = g_2^{d=2}$. For $d=3$, let
\begin{align*}
    g_1^{d=3}(u) &= u_{\epsilon}^{-\frac{3}{2}-\nu} F\left(\tfrac{3}{2}+\nu, \tfrac{1}{2}+\nu; 1+2\nu; u_{\epsilon}^{-1} \right) \, , \\
    g_2^{d=3}(u) &= -\frac{4^{1+\nu}}{\sqrt{\pi}} \frac{\Gamma(1+\nu)}{\Gamma(\frac{3}{2}+\nu)} \, Q'_{\nu-\frac{1}{2}}(2u_{\epsilon}-1) \, .
\end{align*}
Then,
\begin{equation*}
g_1^{d=3}(u) = g_2^{d=3}(u) = u_{\epsilon}^{-\frac{3}{2}-\nu} \left(1 + \frac{\frac{3}{2}+\nu}{2u} + \mathcal{O}(u^{-2}) \right) \, .
\end{equation*}
Hence, $g_1^{d=3} = g_2^{d=3}$.

For arbitrary $d$, we give a proof by induction. For even $d$, let it be true for a fixed $d=2n$. For $d=2(n+1)$, let
\begin{align*}
g_1^{d=2n+2}(u) &= u_{\epsilon}^{-n-1-\nu} F\left(n+1+\nu, \tfrac{1}{2}+\nu; 1+2\nu; u_{\epsilon}^{-1} \right) \, , \\
g_2^{d=2n+2}(u) &= \mathcal{N}^{n+1} \, \frac{\dd^n}{\dd u^n} \frac{\left(2u_{\epsilon}-1+2\sqrt{u_{\epsilon}(u_{\epsilon}-1)}\right)^{-\nu}}{\sqrt{u_{\epsilon}(u_{\epsilon}-1)}} \, ,
\end{align*}
for some constant $\mathcal{N}^{n+1}$. We know that
\begin{equation*}
	g_1^{d=2n}(u) = g_2^{d=2n}(u) = \mathcal{N}^{n} \, \frac{\dd^{n-1}}{\dd u^{n-1}} \frac{\left(2u_{\epsilon}-1+2\sqrt{u_{\epsilon}(u_{\epsilon}-1)}\right)^{-\nu}}{\sqrt{u_{\epsilon}(u_{\epsilon}-1)}} 
\end{equation*}
and that
\begin{align*}
g_2^{d=2n+2}(u) &= \frac{\mathcal{N}^{n+1}}{\mathcal{N}^{n}} \, \frac{\dd}{\dd u} g_2^{d=2n}(u) = \frac{\mathcal{N}^{n+1}}{\mathcal{N}^{n}} \, \frac{\dd}{\dd u} g_1^{d=2n}(u) \\
&= - \frac{\mathcal{N}^{n+1}}{\mathcal{N}^{n}} (n+\nu) u_{\epsilon}^{-n-\nu} \left(1 + \frac{n+1+\nu}{2u} + \mathcal{O}(u^{-2}) \right) \, .
\end{align*}
Comparing with
\begin{align*}
g_1^{d=2n+2}(u) = u_{\epsilon}^{-n-\nu} \left(1 + \frac{n+1+\nu}{2u} + \mathcal{O}(u^{-2}) \right) \, ,
\end{align*}
we conclude that $g_1^{d=2n+2} = g_2^{d=2n+2}$ with 
$$ \mathcal{N}^{n+1} = - \frac{\mathcal{N}^{n}}{n+\nu} = (-1)^n \frac{\mathcal{N}^{1}}{\Gamma(n+1+\nu)} = \frac{(-1)^n \, 4^{\nu}}{\Gamma(n+1+\nu)} \, . $$

Similarly, for odd $d$, let it be true for a fixed $d=2n+1$. For $d=2n+3$, let
\begin{align*}
g_1^{d=2n+3}(u) &= u_{\epsilon}^{-n-\frac{3}{2}-\nu} F\left(n+\tfrac{3}{2}+\nu, \tfrac{1}{2}+\nu; 1+2\nu; u_{\epsilon}^{-1} \right) \, , \\
g_2^{d=2n+3}(u) &= \mathcal{N}^{n+1} \, \frac{\dd^{n+1}}{\dd u^{n+1}} Q_{\nu-\frac{1}{2}}(2u_{\epsilon}-1) \, ,
\end{align*}
for some constant $\mathcal{N}^{n+1}$. We know that
\begin{equation*}
g_1^{d=2n+1}(u) = g_2^{d=2n+1}(u) = \mathcal{N}^{n} \, \frac{\dd^{n}}{\dd u^{n}} Q_{\nu-\frac{1}{2}}(2u_{\epsilon}-1)
\end{equation*}
and that
\begin{align*}
g_2^{d=2n+3}(u) &= \frac{\mathcal{N}^{n+1}}{\mathcal{N}^{n}} \, \frac{\dd}{\dd u} g_2^{d=2n+1}(u) = \frac{\mathcal{N}^{n+1}}{\mathcal{N}^{n}} \, \frac{\dd}{\dd u} g_1^{d=2n+1}(u) \\
&= - \frac{\mathcal{N}^{n+1}}{\mathcal{N}^{n}} \left(n + \frac{1}{2}+\nu\right) u_{\epsilon}^{-n-\frac{3}{2}-\nu} \left(1 + \frac{n+\frac{3}{2}+\nu}{2u} + \mathcal{O}(u^{-2}) \right) \, .
\end{align*}
Comparing with
\begin{align*}
g_1^{d=2n+3}(u) = u_{\epsilon}^{-n-\frac{3}{2}-\nu} \left(1 + \frac{n+\frac{3}{2}+\nu}{2u} + \mathcal{O}(u^{-2}) \right) \, ,
\end{align*}
we conclude that $g_1^{d=2n+3} = g_2^{d=2n+3}$ with 
$$ \mathcal{N}^{n+1} = - \frac{\mathcal{N}^{n}}{n+\nu} = (-1)^n \frac{\mathcal{N}^{1}}{\Gamma(n+\frac{3}{2}+\nu)}
= \frac{4^{1+\nu}}{\sqrt{\pi}} \frac{(-1)^{n+1} \, \Gamma(1+\nu)}{\Gamma(\frac{3}{2}+\nu)\Gamma(n+\frac{3}{2}+\nu)} \, . $$
This concludes the proof.


\section{Delta distribution representation}
\label{app:deltafunction}

In this appendix, we prove the identities
\begin{equation}
\int_0^{\infty} \dd k \, k \, \left(\frac{k}{r}\right)^{\frac{d-3}{2}} \! J_{\frac{d-3}{2}}(kr) 
= 2^{\frac{d-3}{2}} \, \Gamma\left(\frac{d-1}{2}\right) \, \frac{\delta(r)}{r^{d-2}} = \frac{(2\pi)^{\frac{d}{2}}\Gamma\left(\frac{d-1}{2}\right)}{\sqrt{2} \, \Gamma\left(\frac{d}{2}\right)} \prod_{i=1}^{d-1} \delta(x^i-{x'}^i) \, ,
\end{equation}
where $d \geq 2$ is an integer and $r>0$.

We start with a standard representation of the delta distribution (Eq.~(1.17.13) of \cite{NIST}),
\begin{equation*}
\delta(r-r') = r \int_0^{\infty} \dd k \, k \, J_{\mu}(kr) J_{\mu}(kr') \, ,
\end{equation*}
with ${\rm Re}(\mu) > -1$ and $r, r' > 0$. Given that $\delta(r-r')$ is zero when $r \neq r'$, we may write
\begin{equation*}
\delta(r-r') = \frac{r^{\mu+1}}{r'^{\mu}} \int_0^{\infty} \dd k \, k \, J_{\mu}(kr) J_{\mu}(kr') \, .
\end{equation*}
Using
\begin{equation*}
J_{\mu}(kr') = \frac{\left(\frac{1}{2}kr'\right)^{\mu}}{\Gamma(\mu+1)} + \mathcal{O}({r'}^{\mu+1}) \, ,
\end{equation*}
and letting $r' \to 0$, we get
\begin{equation*}
\delta(r) = \frac{r^{\mu+1}}{2^{\mu} \Gamma(\mu+1)} \int_0^{\infty} \dd k \, k^{\mu+1} \, J_{\mu}(kr) \, .
\end{equation*}
Letting $\mu = \frac{d-3}{2}$, this allows us to obtain
\begin{equation*}
\int_0^{\infty} \dd k \, k \, \left(\frac{k}{r}\right)^{\frac{d-3}{2}} \! J_{\frac{d-3}{2}}(kr) 
= 2^{\frac{d-3}{2}} \, \Gamma\left(\frac{d-1}{2}\right) \, \frac{\delta(r)}{r^{d-2}} \, .
\end{equation*}

Finally, by changing from the Cartesian coordinates $x^i$, $i=1, \ldots, d-1$, to spherical coordinates,
\begin{equation*}
\prod_{i=1}^{d-1} \delta(x^i-{x'}^i) = \frac{\delta(r)}{A_{d-1} r^{d-2}}
= \frac{\Gamma\left(\frac{d}{2}\right)}{2\pi^{\frac{d}{2}}} \frac{\delta(r)}{r^{d-2}} \, ,
\end{equation*}
where $A_{d-1}$ is the area of a $(d-1)$-sphere. Hence, we obtain the desired identity
\begin{equation*}
\int_0^{\infty} \dd k \, k \, \left(\frac{k}{r}\right)^{\frac{d-3}{2}} \! J_{\frac{d-3}{2}}(kr) 
= \frac{(2\pi)^{\frac{d}{2}}\Gamma\left(\frac{d-1}{2}\right)}{\sqrt{2} \, \Gamma\left(\frac{d}{2}\right)} \prod_{i=1}^{d-1} \delta(x^i-{x'}^i) \, .
\end{equation*}
%


\section{Eigenfunction expansion of the delta distribution}
\label{app:eigenfunctionexpansion}

In this appendix, we show how to compute the expansion of the Dirac delta distribution in terms of the eigenfunctions of the operator $L$ defined in \eqref{eq:STeq} in an efficient way, as presented e.g.~in Chapter 7 of Ref.~\cite{Stakgold}. These expansions can be found in Section~4.11 of \cite{Titchmarsh}, but the computation presented there involves convoluted and old-fashioned methods.

We present the computation of the expansion \eqref{eq:deltaexpansionnu01} in terms of the eigenfunctions of $L$ which satisfy Robin boundary conditions when $\nu \in (0,1)$. The others can be obtained in a similar way.

First, we compute the \emph{Green's function} $\mathcal{G}(z,z';\lambda)$ associated with the Sturm-Liouville problem \eqref{eq:STeq}, which satisfies
\begin{equation*}
(L\otimes\mathbb{I} - \lambda) \, \mathcal{G} = (\mathbb{I}\otimes L - \lambda) \, \mathcal{G} = \delta(z-z') \, ,
\end{equation*}
and appropriate boundary conditions at $z=0$ and $z'=0$, if necessary. This can be done as follows. For $z < z'$, $\mathcal{G}(z,z';\lambda)$ is the solution of the homogeneous equation in the first entry, $u(z;\lambda)$, satisfying the boundary condition at $z=0$, whereas for $z > z'$, $\mathcal{G}(z,z';\lambda)$ is the solution of the homogeneous equation, $v(z;\lambda)$, which is $L^2(z_0,\infty)$ for some $z_0>0$ and for some $\lambda \in \bC$. Then, ensuring continuity at $z=z'$, one has
\begin{equation*}
\mathcal{G}(z,z';\lambda) = \mathcal{N}_{\lambda} \, u(z_<;\lambda) \, v(z_>;\lambda) \, ,
\end{equation*}
where $z_< \doteq \min\{z,z'\}$ and $z_> \doteq \max\{z,z'\}$. The jump condition,
\begin{equation*}
\left.\frac{\dd}{\dd z}\mathcal{G}(z,z';\lambda)\right|_{z=z'^+} - \left.\frac{\dd}{\dd z}\mathcal{G}(z,z';\lambda)\right|_{z=z'^-} = -1 \, ,
\end{equation*}
fixes the normalization constant
\begin{equation*}
\mathcal{N}_{\lambda} = - \frac{1}{W_z \big[u(\cdot;\lambda), v(\cdot;\lambda)\big]} \, .
\end{equation*}

The Green's function can also be obtained as an expansion in terms of the eigenfunctions of $L$ which satisfy the same boundary conditions. If the operator $L$ only had a point spectrum with real eigenvalues $\lambda_n$ and corresponding eigenfunctions $\psi_n$, it is easy to show that $\mathcal{G}(z,z';\lambda)$ would be written as
\begin{equation*}
\mathcal{G}(z,z';\lambda) = - \sum_n \frac{\psi_n(z) \overline{\psi}_n(z')}{\lambda-\lambda_n} \, .
\end{equation*}
As a function of the complex parameter $\lambda$, $\mathcal{G}$ has simple poles at $\lambda = \lambda_n$ and corresponding residues $-\psi_n(z) \overline{\psi}_n(z')$. Hence, one can write
\begin{equation*}
- \frac{1}{2\pi i} \int_{C_{\infty}} \dd \lambda \, \mathcal{G}(z,z';\lambda) = \sum_n \psi_n(z) \overline{\psi}_n(z') = \delta(z-z') \, ,
\end{equation*}
where $C_{\infty}$ is an infinitely large circle in the $\lambda$ plane and the integral is taken counterclockwise. If there is also a continuous spectrum (as it happens in our case), the Green's function has a branch cut and the integral above, besides the sum of the residues at the eigenvalues, includes a branch-cut integral over a portion of the real axis,
\begin{equation} \label{eq:deltarepintG}
- \frac{1}{2\pi i} \int_{C_{\infty}} \dd \lambda \, \mathcal{G}(z,z';\lambda) = \sum_n \psi_n(z) \overline{\psi}_n(z') + \int \dd \lambda \, \psi_{\lambda}(z) \overline{\psi}_{\lambda}(z')
= \delta(z-z') \, .
\end{equation}
Eq.~\eqref{eq:deltarepintG} allows us to obtain the expansion of the delta distribution in terms of the eigenfunctions of $L$ by performing the integral of the Green's function $\mathcal{G}$, which we obtained above, over the spectral parameter $\lambda$.

In the case at hand, we obtain the expansion in terms of eigenfunctions of the operator $L$ defined in \eqref{eq:STeq} when $\nu \in (0,1)$, satisfying the boundary conditions \eqref{eq:BCsing}. The solution of the homogeneous equation satisfying the boundary condition at $z=0$ may be written as $u(z;\lambda) = \sqrt{z} \, \big[ c J_{\nu}\big(\sqrt{\lambda} z\big) - \lambda^{\nu} J_{-\nu}\big(\sqrt{\lambda} z\big) \big]$, whereas $v(z;\lambda) = \sqrt{z} \, H_{\nu}^{(1)}\big(\sqrt{\lambda} z\big)$ is in $L^2(z_0,\infty)$ for any $z_0>0$ if $\lambda \notin [0,\infty)$. Thus, the Green's function is given by
\begin{equation*}
\mathcal{G}(z,z';\lambda) = -\frac{i\pi}{2} \frac{\sqrt{z_<} \, \Big[ c J_{\nu}\big(\sqrt{\lambda} z_<\big) - \lambda^{\nu} J_{-\nu}\big(\sqrt{\lambda} z_<\big) \Big] \sqrt{z_>} \, H_{\nu}^{(1)}\big(\sqrt{\lambda} z_>\big)}{c- e^{-i\pi\nu} \lambda^{\nu}} \, ,
\end{equation*}
with $\lambda \notin [0,\infty)$ for all $c \in \bR$ and additionally with $\lambda \neq -c^{1/\nu}$ if $c>0$, which is a pole of $\mathcal{G}$. This is the negative eigenvalue in the spectrum with corresponding ``bound state'' eigenfunction of the form $\sqrt{z} \, K_{\nu} \big(c^{1/(2\nu)} z\big)$.

Consider first the case $c<0$, for which there is no point spectrum and the continuous spectrum is $[0,\infty)$. Then, 
\begin{align*}
\delta (z-z') &= - \frac{1}{2\pi i} \int_{C_{\infty}} \dd \lambda \, \mathcal{G}(z,z';\lambda) \\
&= \frac{1}{2\pi i} \int_0^{\infty} \dd |\lambda| \lim_{\epsilon \to 0^+} \big[\mathcal{G}(z,z';|\lambda|+i\epsilon) - \mathcal{G}(z,z';|\lambda|-i\epsilon) \big] \\
&= \sqrt{zz'} \int_0^{\infty} \dd |\lambda| \, \frac{\Big[ c J_{\nu}\big(\sqrt{|\lambda|} z\big) - |\lambda|^{\nu} J_{-\nu}\big(\sqrt{|\lambda|} z\big) \Big] \Big[ c J_{\nu}\big(\sqrt{|\lambda|} z'\big) - |\lambda|^{\nu} J_{-\nu}\big(\sqrt{|\lambda|} z'\big) \Big]}{c^2 - 2c |\lambda|^{\nu} \cos(\pi \nu) + |\lambda|^{2\nu}} \\
&= \sqrt{zz'} \int_0^{\infty} \dd q \, q \, \frac{\left[ c J_{\nu}(qz) - q^{2\nu} J_{-\nu}(qz) \right] \left[ c J_{\nu}(qz') - q^{2\nu} J_{-\nu}(qz') \right]}{c^2 - 2c q^{2\nu} \cos(\pi \nu) + q^{4\nu}} \, ,
\end{align*}
which is Eq.~\eqref{eq:deltaexpansionnu01}. 

Finally, for $c>0$ besides the continuous spectrum $[0,\infty)$ there is the eigenvalue $-c^{1/\nu}$, hence, according to \eqref{eq:deltarepintG}, one adds an extra term,
\begin{align*}
\delta (z-z') &= \sqrt{zz'} \int_0^{\infty} \dd q \, q \, \frac{\left[ c J_{\nu}(qz) - q^{2\nu} J_{-\nu}(qz) \right] \left[ c J_{\nu}(qz') - q^{2\nu} J_{-\nu}(qz') \right]}{c^2 - 2c q^{2\nu} \cos(\pi \nu) + q^{4\nu}} \notag \\
&\quad + 2\sqrt{zz'} \, c^{1/\nu} \, \frac{\sin(\pi \nu)}{\pi \nu} K_{\nu}\big(c^{1/(2\nu)}z\big) K_{\nu}\big(c^{1/(2\nu)}z'\big) \, .
\end{align*}
%



\begin{thebibliography}{999}
	
	\bibitem{Benini:2013fia}
	M.~Benini, C.~Dappiaggi and T.~-P.~Hack,
	Int.\ J.\ Mod.\ Phys.\ A {\bf 28} 1330023 (2013)
	[arXiv:1306.0527 [gr-qc]].
	
	\bibitem{Brunetti:2015vmh}
	R.~Brunetti, C.~Dappiaggi, K.~Fredenhagen and J.~Yngvason,
	{\it Advances in Algebraic Quantum Field Theory}, 
	Springer Verlag (2015).
	
	\bibitem{Kay:1988mu}
	B.~S.~Kay and R.~M.~Wald,
	Phys.\ Rept.\  {\bf 207} (1991) 49.
	
	\bibitem{Khavkine:2014mta} 
	I.~Khavkine and V.~Moretti,
	arXiv:1412.5945 [math-ph].
	
	\bibitem{Allen} 
	B.~Allen, 
	Phys. Rev. {\bf D 32} (1985) 3136.
	
	\bibitem{BD} 
	T.~S.~Bunch and P.~C.~W.~Davies,
	Proc. R. Soc. Lond. A {\bf 360} (1978) 117.
	
	\bibitem{Dappiaggi:2009fx}
	C.~Dappiaggi, V.~Moretti and N.~Pinamonti,
	Adv.\ Theor.\ Math.\ Phys.\  {\bf 15} (2011) no.2,  355
	[arXiv:0907.1034 [gr-qc]].
	
	\bibitem{Gerard}
	C.~G\'erard,
	arXiv:1608.06739 [math-ph].
	
	\bibitem{Sanders:2013vza}
	K.~Sanders,
	Lett.\ Math.\ Phys.\  {\bf 105} (2015) no.4,  575
	[arXiv:1310.5537 [gr-qc]].
	
	\bibitem{Them:2013uka}
	K.~Them and M.~Brum,
	Class.\ Quant.\ Grav.\  {\bf 30} (2013) 235035
	[arXiv:1302.3174 [gr-qc]].
	
	\bibitem{Dappiaggi:2007mx}
	C.~Dappiaggi, V.~Moretti and N.~Pinamonti,
	Commun.\ Math.\ Phys.\  {\bf 285} (2009) 1129
	[arXiv:0712.1770 [gr-qc]].
	
	\bibitem{Dappiaggi:2008dk}
	C.~Dappiaggi, V.~Moretti and N.~Pinamonti,
	J.\ Math.\ Phys.\  {\bf 50} (2009) 062304
	[arXiv:0812.4033 [gr-qc]].
	
	\bibitem{Olbermann:2007gn}
	H.~Olbermann,
	Class.\ Quant.\ Grav.\  {\bf 24} (2007) 5011
	[arXiv:0704.2986 [gr-qc]].
	
	\bibitem{Dappiaggi:2014gea}
	C.~Dappiaggi, G.~Nosari and N.~Pinamonti,
	Math.\ Phys.\ Anal.\ Geom.\  {\bf 19} (2016) no.2,  12
	arXiv:1412.1409 [math-ph].
	
	\bibitem{HawkingEllis}
	S.~Hawking and G.~Ellis,
	{\it The Large Scale Structure of Space-time},
	Cambridge University Press (1973).
	
	\bibitem{Ammon:2015wua}
	M.~Ammon and J.~Erdmenger,
	{\it Gauge/gravity Duality: Foundations and Applications},
	Cambridge University Press (2015).
	
	\bibitem{Kay:1992es}
	B.~S.~Kay,
	Rev.\ Math.\ Phys.\ SI {\bf 1} (1992) 167.  
	
	\bibitem{Avis:1977yn}
	S.~J.~Avis, C.~J.~Isham and D.~Storey,
	Phys.\ Rev.\ D {\bf 18} (1978) 3565.
	
	\bibitem{Allen:1985wd}
	B.~Allen and T.~Jacobson,
	Commun.\ Math.\ Phys.\  {\bf 103}, 669 (1986).
	
	\bibitem{Burges:1985qq}
	C.~J.~C.~Burges, D.~Z.~Freedman, S.~Davis and G.~W.~Gibbons,
	Annals Phys.\  {\bf 167}, 285 (1986).
	
	\bibitem{Kent:2014nya} 
	C.~Kent and E.~Winstanley,
	Phys.\ Rev.\ D {\bf 91}, no. 4, 044044 (2015)
	[arXiv:1408.6738 [gr-qc]].
	
	\bibitem{Belokogne:2016dvd} 
	A.~Belokogne, A.~Folacci and J.~Queva,
	arXiv:1610.00244 [gr-qc].
	
	\bibitem{Ishibashi:2004wx}
	A.~Ishibashi and R.~M.~Wald,
	Class.\ Quant.\ Grav.\  {\bf 21} (2004) 2981
	[hep-th/0402184].
	
	\bibitem{Zahn:2015due}
	J.~Zahn,
	arXiv:1512.05512 [math-ph]. 
	
	\bibitem{Kent:2013}
	C.~Kent,  
	{\it Quantum scalar field theory on anti-de Sitter space,} 
	PhD thesis (2013), University of Sheffield.  
	
	\bibitem{Wald}
	R.~M.~Wald,
	{\it General Relativity}, 1st edn. 
	The University of Chicago Press, (1984). 
	
	\bibitem{Zettl:2005}
	A.~Zettl,
	{\it Sturm-Liouville Theory,}
	American Mathematical Society, (2005).  
	
	\bibitem{Breitenlohner:1982jf} 
	P.~Breitenlohner and D.~Z.~Freedman,
	Annals Phys.\  {\bf 144}, 249 (1982).
	
	\bibitem{Weyl}
	H.~Weyl, 
	Math. Annalen \textbf{68} (1910) 220-269.
	
	\bibitem{Titchmarsh}
	E.~C.~Titchmarsh,
	{\it Eigenfunction Expansions}, part I, 2nd ed., 
	Oxford University Press, (1962).
	
	\bibitem{Danielsson:1998wt}
	U.~H.~Danielsson, E.~Keski-Vakkuri and M.~Kruczenski,
	JHEP {\bf 9901} (1999) 002
	[hep-th/9812007].
	
	\bibitem{Fulling:1987}
	S.~A.~Fulling and S.~N.~M.~Ruijsenaars, 
	Phys.\ Rept.\ {\bf 152} (1989) no. 3 135-176.
	
	\bibitem{Kirsten:1993ug}
	K.~Kirsten and J.~Garriga,
	Phys.\ Rev.\ D {\bf 48} (1993) 567
	[gr-qc/9305013].
	
	\bibitem{Lebedev:1972}
	N.~N.~Lebedev,
	{\it Special Functions \& their Applications,}
	Dover Publications (1972).
	
	\bibitem{Decanini:2005eg} 
	Y.~Decanini and A.~Folacci,
	Phys.\ Rev.\ D {\bf 78}, 044025 (2008)
	[gr-qc/0512118].
	
	\bibitem{Radzikowski:1996pa} 
	M.~J.~Radzikowski,
	Commun.\ Math.\ Phys.\ {\bf 179}, 529 (1996).
	
	\bibitem{Radzikowski:1996ei} 
	M.~J.~Radzikowski,
	Commun.\ Math.\ Phys.\  {\bf 180}, 1 (1996).
	
	\bibitem{DappiaggiFerreira} 
	C.~Dappiaggi, H.~R.~C.~Ferreira,
	to appear.
	
	\bibitem{Vasy}
	A.~Vasy,
	Anal. PDE 5 (2012), no. 1, 81-144. 
	
	\bibitem{Brunetti:2005pr}
	R.~Brunetti, K.~Fredenhagen and S.~Hollands,
	JHEP {\bf 0505} (2005) 063
	[hep-th/0503022]. 
	
	\bibitem{Gradshteyn}
	I.~S.~Gradshteyn, I.~M.~Ryzhik, 
	{\it Table of Integrals, Series and Products}, 7th edn. Academic Press, (2007). 
	
	\bibitem{NIST}
	F.~Olver,
	{\it NIST Handbook of Mathematical Functions,}
	Cambridge University Press (2010).
	
	\bibitem{Stakgold}
	I.~Stakgold and M.~Holst,
	{\it Green's Functions and Boundary Value Problems,}
	3rd ed., John Wiley \& Sons, (2011). 
	
\end{thebibliography}
\end{document}